\documentclass[11pt, reqno]{amsart}

\usepackage{graphicx,amsmath,amssymb,epsfig,harvard}
\usepackage{colortbl}
\long\def\comment#1{}
\long\def\comment#1{}
\newtheorem{algorithm}{Algorithm}
\newtheorem{theorem}{Theorem}

\theoremstyle{definition}

\newtheorem{remark}{Comment}[section]

\newcommand{\citen}{\citeasnoun}

\newcommand{\be}{\begin{eqnarray}}
\newcommand{\ee}{\end{eqnarray}}

\newcommand{\ba}{\begin{array}}
\newcommand{\ea}{\end{array}}
\newcommand{\bs}{\begin{align}\begin{split}\nonumber}
\newcommand{\bsnumber}{\begin{align}\begin{split}}
\newcommand{\es}{\end{split}\end{align}}

\renewcommand{\(}{\left(}
\renewcommand{\)}{\right)}
\renewcommand{\[}{\left[}
\renewcommand{\]}{\right]}
\renewcommand{\hat}{\widehat}

\newcommand{\Ep}{{\mathrm{E}}}
\newcommand{\En}{{\mathbb{E}_n}}

\newcommand{\conflvl}{\gamma}

\newcommand{\PX}{\mathcal{P}_{\hat I}}
\newcommand{\MX}{\mathcal{M}_{\hat I}}
\newcommand{\MXd}{\mathcal{M}_{\hat I_1}}
\newcommand{\MXy}{\mathcal{M}_{\hat I_2}}

\renewcommand{\Pr}{{\mathrm{P}}}

\def\RR{ {\Bbb{R}}}

\def\supp{{\rm support}}

\newcommand{\semin}[1]{\phi_{{\rm min}}(#1)}
\newcommand{\semax}[1]{\phi_{{\rm max}}(#1)}
\renewcommand{\hat}{\widehat}
\renewcommand{\leq}{\leqslant}
\renewcommand{\geq}{\geqslant}

\allowdisplaybreaks

\begin{document}

\title[High-Dimensional Sparse Econometric Models]{Inference for High-Dimensional Sparse Econometric Models}
\author[Belloni \ Chernozhukov \ Hansen]{A. Belloni \and V. Chernozhukov \and C. Hansen}

\date{First version:  June 2010,  This version is of  \today.}

\thanks{The preliminary results of this paper were presented
at V. Chernozhukov's invited lecture at 2010 Econometric Society World
Congress in Shanghai. Financial support
from the National Science Foundation is
gratefully acknowledged.  Computer programs
to replicate the empirical analysis
are available from the authors.  We thank Josh Angrist, the editor Manuel Arellano, the discussant
Stephane Bonhomme, and Denis Chetverikov for
excellent constructive comments that helped us improve the article.}

\maketitle

\begin{abstract}
 This article is about estimation and inference methods for high dimensional sparse (HDS) regression models in econometrics.  High dimensional sparse models arise in situations where many regressors (or series terms) are available and the regression function is well-approximated by
   a parsimonious, yet unknown set of regressors. The latter condition makes it possible to estimate the entire regression function effectively by searching for approximately the right set of regressors.
We discuss methods for identifying this set of regressors and estimating their coefficients based on $\ell_1$-penalization and describe key theoretical results.  In order to capture
realistic practical situations, we expressly allow for imperfect selection of regressors and study the impact of this imperfect selection on estimation and inference results.  We focus the main part of the article on the
use of HDS models and methods in the instrumental variables model and the partially linear model.
We present a set of novel inference results for these models and illustrate their use with applications to returns to schooling and growth regression. \\

\end{abstract}

\section{Introduction}

We consider linear, high dimensional sparse (HDS) regression models in econometrics.  The HDS regression model allows for a large number of regressors, $p$, which is possibly much larger than the sample size, $n$, but imposes that the model is sparse.  That is, we assume only  $s \ll n$ of  these regressors are important for capturing the main features of the regression function. This assumption makes it possible to estimate HDS models effectively by searching for approximately the right set of regressors. In this article, we review estimation methods for HDS models that make use of $\ell_1$-penalization and then provide a set of novel inference results. We also provide empirical examples that illustrate the potential wide applicability of HDS models and methods in econometrics.

The motivation for considering HDS models comes in part from the wide availability of data sets with many regressors. For example, the American Housing Survey records prices as well as a multitude of features of houses sold; and scanner data-sets record prices and numerous characteristics of products sold at a store or on the internet.
HDS models are also partly motivated by the use of series methods
in econometrics.  Series methods use many constructed or series regressors -- regressors formed as transformation of elementary regressors -- to approximate regression functions.  In these applications, it is important to have parsimonious yet accurate approximation of the regression function. One way to achieve this is to use the data to select a small of number of informative terms from among a very large set of control variables or approximating functions. In this article, we formally discuss doing this selection and estimating the regression function.

We organize the article as follows.  In the next section, we introduce the concepts of sparse and approximately sparse regression models in the canonical context of modeling a conditional mean function
and motivate the use of HDS models via an empirical and  analytical examples.   In Section \ref{Sec:SparseMethods}, we discuss some principal estimation methods and mention extensions
of these methods to applications beyond conditional mean models. We discuss some key estimation results for HDS methods and mention various extensions of these results in Section \ref{Sec:ResultsHDSM}.  We then develop HDS models and methods in instrumental variables models with many instruments in Section \ref{Sec:IV} and a partially linear model with many series terms in Section \ref{Sec:Treatment}, with the main emphasis
given to inference.  Finally, we present two empirical examples which motivate the use of these methods in Section \ref{Sec:Empirical}.

\textbf{Notation.}  We allow
for the models to change with the sample size, i.e. we allow for array asymptotics. In particular we assume that $p=p_n$ grows to infinity as $n$ grows, and $s =s_n$ can also grow with $n$, although we require that $s \log p = o(n)$. Thus, all parameters are
implicitly indexed by the sample size $n$, but we omit the index to simplify
notation. 
We also use the following empirical process notation, $\En[f] = \En[f(z_i)] = \sum_{i=1}^n f(z_i)/n.$  The ${l}_2$-norm is denoted by
$\|\cdot\|$, and the ${l}_0$-norm, $\|\cdot\|_0$, denotes the number of non-zero components of a vector.  We use $\| \cdot \|_{\infty}$ to denote the maximal element of a vector.  
Given a vector $\delta \in \RR^p$, and a set of
indices $T \subset \{1,\ldots,p\}$, we denote by $\delta_T \in \RR^p$ the vector in which $\delta_{Tj} = \delta_j$ if $j\in T$, $\delta_{Tj}=0$ if $j \notin T$. We use the notation $(a)_+ = \max\{a,0\}$, $a \vee b = \max\{ a, b\}$ and $a \wedge b = \min\{ a , b \}$. We also use the notation $a \lesssim b$ to denote $a \leqslant c b$ for some constant $c>0$ that does not depend on $n$; and $a\lesssim_P b$ to denote $a=O_P(b)$. For an event $E$, we say that $E$ wp $\to$ 1 when $E$ occurs with probability approaching one as $n$ grows. 


\section{Sparse and Approximately Sparse Regression Models}\label{Sec:SparseModel}
In this section we review the modeling foundations for HDS methods and provide motivating examples with emphasis on applications in econometrics.
First, let us consider the following parametric linear regression model:
$$y_i = x_i'\beta_0 + \epsilon_i, \ \ \epsilon_i \sim N(0, \sigma^2), \ \ \beta_0 \in \Bbb{R}^p, \ \ i=1,\ldots,n$$
$$
T = \supp(\beta_0) \text{ has } s \text{ elements where }   s< n,
$$
where $p> n$ is allowed, $T$ is unknown, and regressors $X=[x_1,\ldots, x_n]'$ are fixed.  We assume Gaussian errors to simplify the presentation of the main ideas throughout the article, but note that this assumption can be eliminated without substantially altering the results.  It is clear that simply regressing $y$ on all $p$ available $x$ variables is problematic  when $p$ is large relative to $n$ which motivates consideration of models that impose some regularization on the estimation problem.

The key assumption that allows effective use of this large set of covariates is sparsity of the model of interest.  Sparsity refers to the condition that only $s \ll n$ elements of $\beta_0$ are non-zero but allows the identities of these elements to be unknown.  Sparsity can be motivated on economic grounds in situations where a researcher believes that the economic outcome could be well-predicted by a small (relative to the sample size) number of factors but is unsure about the identity of the relevant factors.    Note that we allow $s=s_n$ to grow with $n$, as mentioned in the notation section, although $s \log p = o(n)$ will be required for consistency. This simple sparse model substantially generalizes the classical parametric linear model by letting the identities, $T,$ of the relevant regressors be unknown.
This generalization is useful in practice since it is problematic to assume that we know the identities of the relevant regressors in many  examples.

The previous model is simple and allows us to convey the essential ideas of the sparsity-based approach.  However, it is unrealistic in that it presumes \emph{exact} sparsity or that, after accounting for $s$ main regressors, the error in approximating the regression function is zero.
We shall make no formal use of the previous model,
but instead use a much more general, \emph{approximately sparse} or nonparametric model.  In this model,
\emph{all} of the regressors potentially have a \emph{non-zero} contribution to the regression function, but no more than
$s$ unknown regressors are needed for approximating the regression function with a sufficient degree
of accuracy.


We formally define the approximately sparse model as follows.

\noindent \textbf{Condition ASM.} \emph{We have data $\{(y_i,z_i), i=1,\ldots,n\}$  that for each $n$ obey the regression model:
\begin{equation}\label{Def:NP}
y_i = f(z_i) + \epsilon_i, \ \   \epsilon_i \sim N(0, \sigma^2),  \ \  i=1,\ldots,n,
\end{equation}
where $y_i$ is the outcome variable, $z_i$ is a $k_z$-vector of elementary regressors,
 $f(z_i)$ is the regression function, and $\epsilon_i$ are i.i.d. disturbances. Let $x_i=P(z_i)$, where $P(z_i)$ is a vector
of dimension $p=p_n$, that contains a dictionary of possibly technical transformations of $z_i$, including a constant. The values $x_1,\ldots, x_n$ are treated fixed, and normalized so that $\En[x_{ij}^2] = 1$ for $j=1,\ldots,p$. The regression function $f(z_i)$ admits the approximately sparse form, namely
there exists $\beta_0$ such that
\begin{equation}\label{Def:ASM}
f(z_i) = x_i'\beta_0 + r_i,  \ \ \| \beta_0\|_0 \leq s, \ \   c_s:= \{\En[r_i^2]\}^{1/2} \leq K \sigma \sqrt{s/n}.
\end{equation}
where $s=s_n=o(n/\log p)$ and $K$ is a constant independent of $n$. }

In the set-up we consider the fixed design case, which covers random sampling as a special case where $x_1,\ldots, x_n$ represent a realization of this sample on which we condition throughout.
The vector $x_i=P(z_i)$ can include  polynomial or spline transformations
of the original regressors $z_i$ see, e.g., \citen{newey:series} and \citen{chen:Chapter} for various examples of series terms.
The approximate sparsity can be motivated similarly to \citen{newey:series}, who assumes
that the first $s=s_n$ series terms can approximate the nonparametric regression function well.  Condition ASM
 is more general in that it does not impose that the most important $s=s_n$ terms in the approximating
dictionary are the first $s$ terms; in fact, the identity of the
most important terms is treated as unknown. We note that in the parametric case,  we may naturally choose $x_i'\beta_0=f(z_i)$ so that $r_i=0$ for all $i=1,\ldots,n$. In the nonparametric case, we may think of $x_i'\beta_0$ as any sparse parametric model that yields a good approximation to the true regression function $f(z_i)$ in equation (\ref{Def:NP}) so that $r_i$ is ``small'' relative to
the conjectured size of the estimation error. Given (\ref{Def:ASM}), our target in estimation is the parametric function $x_i'\beta_0$, where we can
call
$$
T := \supp(\beta_0)
$$
the ``true" model. Here we emphasize that the ultimate target in estimation is, of course, $f(z_i)$. The function
$x_i'\beta_0$ is simply a convenient intermediate target introduced so that we can approach the estimation problem as if it were parametric. Indeed, the two targets, $f(z_i)$ and $x_i'\beta_0$, are equal up to the approximation error $r_i$. Thus, the problem of estimating the parametric target $x_i'\beta_0$ is equivalent to
the problem of estimating  the  nonparametric target $f(z_i)$
modulo approximation errors.

One  way to explicitly construct a good approximating model $\beta_0$ for (\ref{Def:ASM}) is by taking $\beta_0$
as the solution to \begin{equation}\label{oracle}
 \min_{  \beta \in \RR^p }  \En [(f(z_i) - x_i'\beta)^2] + \sigma^2 \frac{\|\beta\|_0}{n}.
\end{equation}
We can call (\ref{oracle}) the oracle problem,\footnote{By definition the oracle knows the risk function of any estimator, so it can
compute the best sparse least square estimator.  Under some mild condition the problem of minimizing prediction risk
amongst all sparse least square estimators is equivalent to the problem written here; see, e.g., \citen{BC-LectureNotes}.} and  so we can call
$T = \supp(\beta_0)$ the oracle model.  Note that we necessarily have that $s=\|\beta_0\|  \leqslant n$. The oracle problem (\ref{oracle}) balances
the approximation error $\En [(f(z_i) - x_i'\beta)^2]$ over the design points with the variance term $\sigma^2 \|\beta\|_0/n$, where the latter is determined by the
number of non-zero coefficients in $\beta$.  Letting $ c^2_s:= \En[r^2_i] =  \En [(f(z_i) - x_i'\beta_0)^2]$
denote the squared error from approximating values $f(z_i)$ by $x_i'\beta_0$, the quantity $ c^2_s +  \sigma^2 s/n$ is  the optimal value of (\ref{oracle}).
In common nonparametric problems, such as the one described below, the
optimal solution in (\ref{oracle}) would balance the approximation error  with the variance term giving that  $c_s \leqslant K \sigma \sqrt{s/n}.$
Thus, we would have $ \sqrt{c^2_s +  \sigma^2 s/n} \lesssim \sigma \sqrt{s/n},$ implying that the quantity $\sigma \sqrt{s/n}$ is the ideal goal for the rate of convergence. If we knew the oracle  model $T$, we would achieve this rate by using the oracle estimator, the least squares estimator based on this model.  Of course, we do not generally know $T$ since we do not observe the $f(z_i)$'s and thus cannot attempt to solve the oracle problem (\ref{oracle}).  Since
$T$ is unknown, we will not generally be able to achieve the exact oracle rates of convergence, but we can hope to come close to this rate.

Before considering estimation methods, a natural question is whether exact or approximate HDS models make sense in econometric applications.
In order to answer this question, it is helpful to consider the following two examples in which we abstract from estimation completely and only
ask whether it is possible to accurately describe some structural econometric function $f(z)$ using a low-dimensional approximation
of the form $P(z)'\beta_0$.

\textbf{Example 1: Sparse Models for Earning Regressions}.   In this example  we consider a model for the conditional
expectation of log-wage $y_i$ given education $z_i$, measured in years of schooling.
We can expand  the conditional expectation of wage $y_i$ given education $z_i$:
\begin{equation}\label{general}
E[y_i|z_i] = \sum_{j=1}^p
\beta_{0j} P_j(z_i),
 \end{equation}
 using some dictionary of approximating functions $P(z_i) = (P_1(z_i),\ldots, P_p(z_i))'$, such as polynomial or spline transformations in $z_i$ and/or indicator variables for levels of
 $z_i$. In fact, since we can consider an overcomplete dictionary, the representation of the function using $P_1(z_i),\ldots, P_p(z_i)$ may not be unique, but this is not important for our purposes.

A conventional sparse approximation employed in econometrics is, for example,
\begin{equation}\label{conventional}
f(z_i):=E[y_i|z_i] = \tilde \beta_1P_1(z_i) + \cdots + \tilde \beta_s P_s(z_i)  + \tilde r_i,
 \end{equation}
where the $P_j$'s are low-order polynomials or splines, with typically one or two (linear or linear and quadratic) terms.  Of course, there is no guarantee that the approximation error $\tilde r_i$ in this case is small or that these particular polynomials form the best possible $s$-dimensional approximation. Indeed, we might expect  the function
$E[y_i|z_i]$ to change rapidly near the schooling levels associated with advanced degrees, such as MBAs or MDs.  Low-degree polynomials may not be able to capture this behavior very well, resulting in large approximation errors $\tilde r_i$.

A sensible question is then, ``Can we  find a
better approximation that uses the same number of parameters?'' More formally,  can we construct a much better approximation of the sparse form
\begin{equation}\label{eq:sparse}
f(z_i):=E[y_i|z_i] =  \beta_{k_1}P_{k_1}(z_i) + \cdots + \beta_{k_s} P_{k_s}(z_i) + r_i,
\end{equation}
for some regressor indices $k_1,\ldots,k_s$ selected from $\{1,\ldots,p\}$? Since we can
always include (\ref{conventional}) as a special case, we can in principle do no worse
than the conventional approximation; and, in fact, we can construct (\ref{eq:sparse})  that is much better, if there are some important higher-order terms in (\ref{general}) that are completely missed by the conventional approximation.  Thus, the answer to the question depends strongly on the empirical context.

\begin{figure}[!h]
\centering
\includegraphics[width=0.7\textwidth]{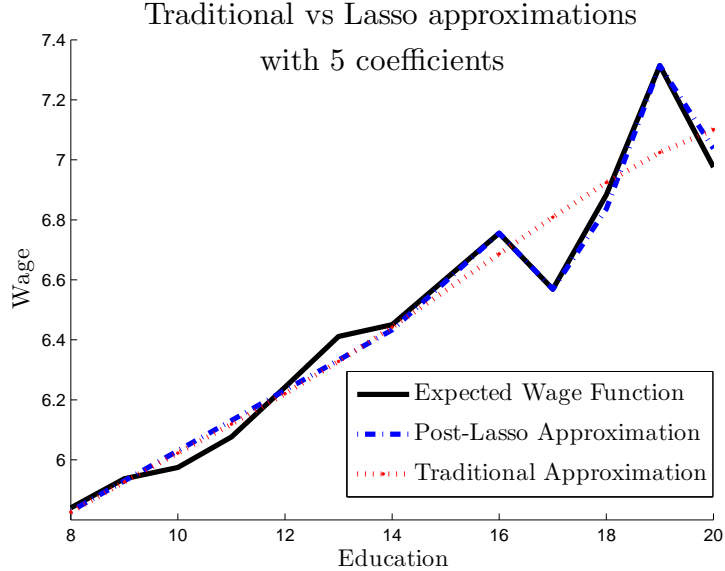} 
\caption{The figures illustrates the Post-Lasso sparse approximation and the fourth order polynomial approximation of the wage function.}\label{Fig:Wage}
\end{figure}

Consider for example the earnings of prime age white males in the 2000 U.S. Census see, e.g., \citen{ACF2006}. Treating this data as the population data, we can compute $f(z_i)=E[y_i|z_i]$ without error. Figure \ref{Fig:Wage} plots this function.   We then construct two sparse approximations and also plot them in Figure \ref{Fig:Wage}. The first is the conventional approximation of the form (\ref{conventional}) with $P_1, \ldots, P_s$ representing polynomials of degree zero to $s-1$ ($s=5$ in this example). The second is an approximation of the form (\ref{eq:sparse}), with $P_{k_1}$, \ldots, $P_{k_s}$ consisting of a constant, a linear term, and three linear splines terms with knots located at 16, 17, and 19 years of schooling.  We find the latter approximation automatically using the $\ell_1$-penalization or Lasso methods discussed below,\footnote{The set of functions considered consisted of 12 linear splines with various knots and monomials of degree zero to four. Note that there were only 12 different levels of schooling.}
 although in this special case we could construct such an approximation just by eye-balling Figure \ref{Fig:Wage} and noting that most of the function is described by a linear function with a few abrupt changes that can be captured by linear spline terms that induce large changes in slope near 17 and 19 years of schooling.  Note that an exhaustive search for a low-dimensional approximation in principle requires looking at a very large set of models. Methods for HDS models, such as $\ell_1$-penalized least squares (Lasso), which we employed in this example, are designed to avoid this search.  \qed

\begin{center}
\begin{table}
\begin{center}
\begin{tabular}{rcccccc}
\hline \hline Sparse Approximation  & &  & & {   $L_2$ error } & & {  $L_\infty$ error } \\ \hline
Conventional  & &  & & 0.12 &  &   0.29       \\
Lasso  & &  & & 0.08 &  & 0.12             \\
Post-Lasso  & &  & & 0.04 &  &   0.08             \\
 \hline
\hline
\\
\end{tabular}
\end{center}\caption{\footnotesize Errors of Conventional and the Lasso-based Sparse Approximations of the Earning Function. The Lasso method minimizes the least squares criterion plus the $\ell_1$-norm of the coefficients scaled by a penalty parameter $\lambda$. The nature
of the penalty forces many coefficients to zero, producing a sparse fit. The Post-Lasso minimizes the least squares criterion over the non-zero components selected by the Lasso estimator. This example deals with a pure approximation problem, in which there is no noise.}
\end{table}
\end{center}

\textbf{Example 2: Series approximations and Condition ASM.} It is clear
from the statement of Condition ASM that this expansion incorporates both
substantial generalizations and improvements over
 the conventional series approximation of regression functions in \citen{newey:series}.
In order to explain this consider the set $\{P_j(z), j\geq 1\}$ of orthonormal basis functions on $[0,1]^d$, e.g. orthopolynomials, with respect
to the Lebesgue measure. Suppose $z_i$ have a uniform distribution on $[0,1]^d$ for simplicity.\footnote{The discussion in this example
continues to apply when $z_i$ has a density that is bounded from above and away from zero on  $[0,1]^d$.}  Assuming
$\Ep[f^2(z_i)] < \infty$, we can represent $f$ via a Fourier expansion,
$
f(z) = \sum_{j=1}^\infty \delta_j P_j(z),$
where $\{\delta_j, j \geq 1\}$ are Fourier coefficients that satisfy $\sum_{j=1}^\infty \delta_j^2 < \infty$.

Let us consider the case that $f$ is a smooth function so that Fourier coefficients
feature a polynomial decay $\delta_j \propto  j^{-\nu}$, where $\nu$ is a measure of smoothness of $f$. Consider the conventional series expansion that uses the first $K$ terms for approximation,
$f(z) = \sum_{j=1}^K \beta_{0j}P_j(z) + a_c(z)$, with $\beta_{0j} = \delta_j$. Here $a_c(z_i)$ is
the approximation error which obeys $ \sqrt{\En[a^{2}_{c}(z_i)]} \lesssim_P \sqrt{\Ep[a^{2}_{c}(z_i)]} \lesssim K^{\frac{-2\nu+1}{2}}$.
Balancing the order $K^{\frac{-2\nu+1}{2}}$ of approximation error  with the order  $\sqrt{K/n}$ of the estimation error
gives the oracle-rate-optimal number of series terms $s = K \propto n^{1/2\nu}$, and the resulting
oracle series estimator, which knows $s$,
will estimate $f$ at the oracle rate of   $n^{\frac{1-2\nu}{4\nu}}$. This also gives us the identity
of the most important series terms $T_{} = \{1,...,s\}$, which are simply the first $s$ terms.  We conclude that
Condition ASM holds for the sparse approximation
$f(z) = \sum_{j=1}^p \beta_{0j}P_j(z) + a_{}(z)$, with $\beta_{0j} = \delta_j$ for $j \leq s$
and $\beta_{0j} = 0$ for $s+ 1\leq j \leq p$, and $a_{}(z_i)=a_{c}(z_i)$, which coincides with the conventional series approximation above, so that
$\sqrt{\En[a^2_{}(z_i)]} \lesssim_P \sqrt{s/n}$ and  $\|\beta_{0}\|_0 \leq s$.

Next suppose that Fourier coefficients
feature the following pattern $\delta_j = 0$ for $j \leq M$ and $\delta_j \propto (j-M)^{-\nu}$ for $j > M$.
Clearly in this case the standard series approximation based on the first $K \leq M$ terms, $\sum_{j=1}^K \delta_jf_j(z)$, has no predictive power for $f(z)$, and the corresponding standard series estimator based on the first $K$ terms therefore fails completely.\footnote{This is not merely a finite sample
phenomenon but is also accommodated in the asymptotics since we expressly allow for array asymptotics; i.e.
the underlying true model could change with $n$. Recall that we omit the indexing by $n$ for ease
of notation.}
In contrast, Condition ASM is easily satisfied in this case, and the Lasso-based
estimators will perform at a near-oracle level in this case. Indeed,
we can use the first $p$ series terms to form the approximation
$
f(z)= \sum_{j=1}^p \beta_{0j}  P_j(z) + a_{}(z)$, where $
\beta_{0j} = 0$ for $j \leq M$ and $j > M + s$, $\beta_{0j} = \delta_j$  for $M+1 \leq j \leq M +s $ with $s \propto n^{1/2\nu}$, and $p$ such that $M + n^{1/2\nu} = o(p).$
Hence $\|\beta_{0}\|_0 = s$, and we have that
$
\sqrt{\En[a^2_{}(z_i)]} \lesssim_P \sqrt{\Ep[a^2_{}(z_i)]} \lesssim \sqrt{s/n} \lesssim
n^{\frac{1-2\nu}{4\nu}}$. \qed

\section{Sparse Estimation Methods}\label{Sec:SparseMethods}

\subsection{$\ell_1$-penalized and post $\ell_1$-penalized estimation methods}

In order to discuss estimation consider first, as a matter of motivation, the classical AIC/BIC type estimator \cite{Akaike1974,Schwarz1978} that solves the empirical (feasible) analog of the oracle problem:
$$
\min_{\beta \in \Bbb{R}^p} \En[(y_i - x_i'\beta)^2] + \frac{\lambda}{n}  \|\beta\|_{0},
$$
where  $\lambda$ is a penalty level.\footnote{The penalty level $\lambda$ in the AIC/BIC type estimator needs to account for the noise since it observes $y_i$ instead of $f(z_i)$ unlike the oracle problem (\ref{oracle}).}  This estimator has attractive theoretical properties. Unfortunately, it is computationally prohibitive since the solution to the problem may require solving $\sum_{k \leqslant n} \binom{p}{k}$ least squares problems.\footnote{Results on the computational intractability of this problem were established in \citen{Natarajan1995}, \citen{GeJiangYe2010} and \citen{ChenGeWangYe2011}.}

One way to overcome the computational difficulty is to consider a convex relaxation of the preceding problem, namely
to employ a closest convex penalty -- the $\ell_1$ penalty -- in place of the $\ell_0$ penalty.  This construction leads to the so called Lasso
estimator $\widehat \beta$ \cite{T1996}, defined as a solution for the following optimization problem:\begin{equation}\label{Def:LASSOmain}
 \min_{\beta \in \Bbb{R}^p} \En[(y_{i} - x_i'\beta)^2] +  \frac{\lambda}{n} \| \beta \|_{1},
\end{equation}
where $\|\beta\|_{1} = \sum_{j=1}^p | \beta_j|$.  The Lasso estimator is computationally attractive because it minimizes a convex function.
A basic choice for penalty level suggested by \citen{BickelRitovTsybakov2009} is
 \begin{equation}\label{Def:LambdaLASSOboound0}\lambda = 2 \cdot c \sigma  \sqrt{2 n \log(2p/\conflvl)}.
 \end{equation}
where $c>1$  and $1-\gamma$ is a confidence level that needs to be set close to 1. The formal motivation for this penalty is that it leads to near-oracle rates of convergence of the estimator.

The penalty level specified above is not feasible since it depends on the unknown $\sigma$.  \citen{BC-PostLASSO} propose  to set
 \begin{equation}\label{Def:LambdaLASSOboound}\lambda = 2 \cdot c \hat \sigma  \Phi^{-1}(1-\conflvl/2p),
 \end{equation}
with $\hat \sigma = \sigma + o_P(1)$  obtained via an iteration method defined in Appendix A, where $c>1$  and $1-\gamma$ is a confidence level.\footnote{Practical recommendations
include the choice $c=1.1$ and $\gamma=.05$.}
 \citen{BC-PostLASSO} also propose the $X$-dependent penalty level: \begin{equation}\label{Def:LambdaLASSO} \lambda = c \cdot  2\hat \sigma \Lambda(1-\conflvl|X),\end{equation}
where
$$\Lambda(1-\conflvl|X) = (1-\conflvl)-\text{quantile of} \ \ n\|\En[x_ig_i]\|_\infty \mid X$$
where $X=[x_1,\ldots,x_n]'$ and $g_i$ are i.i.d. $N(0,1)$ , which can be easily approximated by simulation.  We note that
 \begin{equation}\label{Eq:GaussIneq}
 \Lambda(1-\conflvl|X) \leq  \sqrt{n}\Phi^{-1}(1-\conflvl/2p) \leq \sqrt{2 n \log(2p/\conflvl)},
 \end{equation}
 so $\sqrt{2 n \log(2p/\conflvl)}$ provides a simple upper bound on the penalty level.
Note also that \citen{BellChenChernHans:nonGauss} formulate a  feasible Lasso procedure for the case with heteroscedastic, non-Gaussian disturbances.
We shall refer to the feasible Lasso method with the feasible penalty levels (\ref{Def:LambdaLASSOboound}) or (\ref{Def:LambdaLASSO}) as the \emph{Iterated Lasso}.
This estimator has statistical performance that is similar to that of the (infeasible) Lasso described above.

\citen{BCW-SqLASSO} propose a variant called the \emph{Square-root Lasso} estimator $\widehat \beta$ defined as a solution to the following program: \begin{equation}\label{Def:SQLASSOmain}
\min_{\beta \in \Bbb{R}^p} \sqrt{\En[(y_{i} - x_i'\beta)^2]} +  \frac{\lambda}{n}  \| \beta \|_{1},
\end{equation}
 with the penalty level
\begin{equation}\label{our penalty}
  \lambda = c \cdot  \widetilde \Lambda(1-\conflvl|X),
\end{equation}
 where $c>1$  and
$$\widetilde \Lambda(1-\conflvl|X)= (1-\conflvl)-\text{quantile of } \ n\|\En[x_ig_i]\|_\infty/\sqrt{\En[g_i^2]} \mid X,$$  with $g_i \sim N(0,1)$ independent for $i=1,\ldots, n$.
As with Lasso, there is also  simple asymptotic option for setting the penalty level:
 \begin{equation}\label{our penalty asymptotic}
 \lambda = c \cdot \Phi^{-1}(1-\conflvl/2p).
 \end{equation}
The main attractive feature of (\ref{Def:SQLASSOmain}) is that  the penalty level $\lambda$ is independent of the value $\sigma$,
  and so it is pivotal with respect to that parameter.
  Nonetheless, this estimator has statistical performance that is similar to that of the (infeasible) Lasso described above.
   Moreover, the estimator is a solution to a highly tractable
  conic programming problem:
   \begin{equation}\label{Def:SQLASSOmainConic}
\min_{t \geq 0, \beta \in \Bbb{R}^p} t +  \frac{\lambda}{n}  \| \beta \|_{1}:  \ \
\sqrt{\En[(y_{i} - x_i'\beta)^2]} \leq t,
\end{equation}
where the criterion function is linear in parameters $t$ and positive and negative components of $\beta$,
while the constraint can be formulated with a second-order cone, informally known also as the ``ice-cream cone".

There are several other estimators that make use of penalization by the $\ell_1$-norm.   An important case includes the Dantzig selector estimator proposed and analyzed by \citen{CandesTao2007}. It also relies on $\ell_1$-regularization but exploits the notion that the residuals should be nearly uncorrelated with the covariates. The estimator is defined as a solution to:\begin{equation}\label{Def:DantzigSelector}
\min_{\beta \in \Bbb{R}^p}  \ \  \| \beta \|_{1} \ \ : \ \ \| \En[x_i(y_i-x_i'\beta)]\|_\infty \leq \lambda/n\end{equation} where $\lambda = \sigma \Lambda(1-\conflvl|X)$.
In what follows we will focus our discussion on Lasso but virtually all theoretical results carry over to other $\ell_1$-regularized estimators including  (\ref{Def:SQLASSOmain}) and (\ref{Def:DantzigSelector}).  We also refer to \citen{gautier:tsybakov} for a feasible Dantzig
estimator that combines the square-root lasso method (\ref{Def:SQLASSOmainConic}) with the Dantzig method.

$\ell_1$-regularized estimators often have a  substantial shrinkage bias. In order to remove some of this bias, we consider the post-model-selection estimator that applies ordinary least squares regression to  the model $\widehat T$ selected by a $\ell_1$-regularized estimator $\hat\beta$. Formally, set
$$\widehat T = \supp( \hat \beta ) = \{ j \in \{1,\ldots,p\} \ : \ |\hat\beta_j| > 0\},$$
and define the post model selection estimator $\widetilde \beta$ as \begin{equation}\label{Def:TwoStep} \widetilde \beta \in \arg\min_{\beta \in \mathbb{R}^p} \ \En[(y_i-x_i'\beta)^2]\ \ :  \ \ \beta_j = 0 \text{ for each } j \in \widehat T^c,
\end{equation}
where $\widehat T^c = \{1,...,p\} \setminus \widehat T$. In  words, the estimator is ordinary least squares applied to the data after removing the regressors that were not selected in $\widehat T$.
When the $\ell_1$-regularized method used to select the model is Lasso (Square-root Lasso), the post-model-selection estimator is called Post-Lasso (Post-Square-root Lasso).   If model selection works perfectly -- that is, $\widehat T = T$ --
then the post-model-selection estimator is simply the oracle estimator whose properties are well-known. However, perfect model selection is unlikely in many situations, so we are interested in the properties of the post-model-selection estimator when model selection is imperfect, i.e. when $\widehat T \neq T$, and are especially interested in cases where $T\nsubseteq \widehat T$. In Section \ref{Sec:ResultsHDSM}
we describe the formal properties of the Post-Lasso estimator.

\subsection{Some Heuristics via Convex Geometry}  Before proceeding to the formal results
on estimation, it is useful to consider some heuristics for the $\ell_1$-penalized estimators
and the choice of the penalty level.  For this purpose we consider a parametric model,
and a generic $\ell_1$-regularized estimator based on a differentiable criterion function $\widehat Q$:
\begin{equation}\label{Def:L1reg} \widehat \beta \in \arg \min_{\beta\in \RR^p} \widehat Q(\beta) + \frac{\lambda}{n}\|\beta\|_1,\end{equation}
where, e.g., $ \widehat Q(\beta)  = \En[(y_i - x_i'\beta)^2]$ for Lasso and $ \widehat Q(\beta)  = \sqrt{\En[(y_i - x_i'\beta)^2]}$ for Square-root Lasso. The key quantity in the analysis of (\ref{Def:L1reg}) is the score -- the gradient of $\widehat Q$ at the true value\footnote{In the case of a nonparametric model the score is similar to the gradient of $\widehat Q$ at $\beta_0$ but ignores the approximation errors $r_i$'s.}:
$$
S =  \nabla \widehat Q(\beta_0).
$$
The score $S$ is  the effective ``noise" in the problem that should be dominated by the regularization. However we would like to make the regularization bias as small as possible.  This reasoning suggests choosing the smallest penalty level $\lambda$ that is large enough to dominate the noise with high probability, say $1 - \gamma$, which yields
\begin{equation}\label{BRT principle}
\lambda > c\Lambda,  \text{ for } \Lambda := n\|
S\|_{\infty}, \ \ \
\end{equation}
 where $\Lambda$ is the
maximal score scaled by $n$, and $c>1$ is a theoretical constant of
\citen{BickelRitovTsybakov2009} that guarantees that the score is dominated.  We note that the principle
of setting $\lambda$ to dominate the score of the criterion
function
is a general principle
that carries over to other convex problems with possibly non-differentiable criterion functions and
that leads to the optimal -- near-oracle -- performance of $\ell_1$-penalized
estimators. See, for instance, \citen{BC-SparseQR}.

It is useful to mention some simple heuristics for the principle (\ref{BRT principle}) which arise from considering the simplest case where none of the regressors are significant so that $\beta_0 =0$.
We want our estimator to perform at a near-oracle level in all cases, including this case, but here the oracle estimator $ \beta^*$ sets $\beta^*= \beta_0 =0$.  We also want $\widehat \beta = \beta_0 = 0$ in this case, at least with a high probability, say $1-\conflvl$.  From the subgradient optimality conditions for (\ref{Def:L1reg}), we must have
$$
- S_j + \lambda/n > 0 \text{ and }  S_j + \lambda/n > 0 \text{ for  all } 1 \leq j \leq p$$
for this to be true.
We can only guarantee this by setting the penalty level $\lambda/n$ such that
$
\lambda > n \max_{1 \leq j\leq p} | S_j| = n \| S\|_{\infty}$ with probability at least $1-\conflvl$.
This is precisely the rule (\ref{BRT principle}) appearing above.

Finally, note that in the case of Lasso and Square-root Lasso we have
the following expressions for the score:
$$ \begin{array}{rl}
{\rm Lasso:} & \displaystyle S = 2\En[x_i\epsilon_i] =_d 2  \sigma \En[x_i g_i],\\
\\
{\rm Square\mbox{-}root \ Lasso:} & \displaystyle S = \frac{\En[x_i\epsilon_i]}{\sqrt{\En[\epsilon_i^2]}}=_d\frac{\En[x_ig_i]}{\sqrt{\En[g_i^2]}},\\
\end{array}$$
where $g_i$ are i.i.d. $N(0,1)$ variables. Note that the score for Square-root Lasso is pivotal, while the score for Lasso is not, as it depends on $\sigma$. Thus, the choice of the penalty level for Square-root Lasso need not depend on $\sigma$ to produce near-oracle performance for this estimator.

\subsection{Beyond Mean Models}

Most of the literature on high dimensional sparse models focuses on the mean regression model discussed above. Here we discuss
methods that have been proposed to deal with quantile regression and generalized linear models in high-dimensional sparse settings.
We assume i.i.d. sampling for $(y_i,x_i)$ in this subsection.

\subsubsection{Quantile Regression}

We consider a response variable $y_i$ and $p$-dimensional covariates $x_i$ such that the $u$-th conditional quantile function of $y_i$ given $x_i$ is given by
 \begin{equation}\label{lin model}
F^{-1}_{y_i|x_i}(u|x)= x'\beta(u),  \ \
\beta(u) \in \RR^p,
 \end{equation} where $u \in (0,1)$ is quantile index of interest. Recall that the $u$-th conditional quantile $F^{-1}_{y_i|x_i}(u|x)$ is the  inverse of the conditional distribution function $F_{y_i|x_i}(y|x)$
 of $y_i$ given $x_i=x$. Suppose that the true model $\beta(u)$ has a sparse support: $$T_u = \supp(\beta(u)) = \{ j
\in \{1,\ldots,p\} \ : \ |\beta_j(u)|>0 \}
$$ has only $s_u \leq s \leq n/\log(n \vee p)$ non-zero components.

The population coefficient $\beta(u)$ is known to be a minimizer of the criterion function
 \be\label{define beta}
Q_u(\beta) = \Ep[\rho_{u} (y_i-x_i'\beta)],
 \ee where $\rho_{u} (t) = (u - 1\{t\leq 0\})t$ is the asymmetric absolute deviation  function; see \citen{Koenker:1978}.  Given a random sample  $(y_1,x_1),\ldots,(y_n,x_n)$, $\hat\beta(u)$, the quantile regression estimator of $\beta(u)$, is defined as a minimizer of the empirical analog of (\ref{define beta}):
\begin{equation}\label{QR}
\hat Q_u(\beta) = \En\[ \rho_u (y_i - x_i'\beta) \].
\end{equation}
As before, in high-dimensional settings, ordinary quantile regression is generally not consistent, which motivates the use of penalization in order to remove all, or at least nearly all, regressors whose population coefficients are zero.  The $\ell_1$-penalized quantile regression estimator $\hat \beta (u)$ is a solution to the following optimization problem: \begin{equation}\label{Def:L1QR}
\min_{\beta \in \mathbb{R}^p} \ \hat Q_u(\beta) + \frac{ \lambda\sqrt{u(1-u)}}{n} \|\beta\|_1.
\end{equation} The criterion function in (\ref{Def:L1QR}) is the sum of the criterion function (\ref{QR}) and a penalty function given by a scaled $\ell_1$-norm of the parameter vector.

In order to describe choice of the penalty level $\lambda$, we
introduce the random variable \begin{equation}\label{almost sure}
\Lambda = n \max_{1\leq j\leq p}\left|
\En\[ \frac{x_{ij}( u - 1\{u_i \leq u
\})}{ \sqrt{u(1-u)}}\] \right| ,
\end{equation}
where $u_1,\ldots, u_n$ are i.i.d. uniform $(0,1)$ random
variables, independently distributed from the regressors,
$x_1,\ldots, x_n$.  The random variable $\Lambda$ has a pivotal distribution conditional on  $X= [x_1,\ldots, x_n]'$.
Then, for $c>1$, \citen{BC-SparseQR} propose to set
\begin{equation}\label{Def:LambdaPivotal00}
\lambda = c \cdot  \Lambda(1-\gamma|X),  \text{ where } \
\Lambda(1-\gamma|X) := \textrm{$(1-\gamma)$-quantile of } \Lambda
\text{ conditional  on } X,
 \end{equation}
and $1-\gamma$ is a confidence level that needs to be set close to 1.

The post-penalized estimator (post-$\ell_1$-QR) applies ordinary quantile regression to  the model $\widehat T_u$ selected by the $\ell_1$-penalized quantile regression \cite{BC-SparseQR}. Specifically, set
$$\widehat T_u = \supp( \hat \beta(u) ) = \{ j \in \{1,\ldots,p\} \ : \ |\hat\beta_j(u)| > 0\},$$
and define the post-penalized estimator $\widetilde \beta(u)$ as \begin{equation}\label{Def:TwoStep}
\widetilde \beta(u) \in \arg\min_{\beta \in \mathbb{R}^p}\widehat Q_u(\beta) \ : \ \beta_{j} = 0, \ \ j \in \widehat T_u^c
\end{equation}
which is just ordinary quantile regression removing the regressors that were not selected
in the first step.   \citen{BC-SparseQR} derive the basic properties of the estimators above;
see also \citen{kato} for further important results in nonparametric setting, where group penalization is also
studied.

\subsubsection{Generalized Linear Models}

From the discussion above, it is clear that $\ell_1$-regularized methods can be extended to other criterion functions $\widehat Q$ beyond least squares and quantile regression. $\ell_1$-regularized generalized linear models were considered in \citen{vdGeer}. Let $y \in \RR$ denote  the response variable and $x \in \RR^p$ the covariates. The criterion function of interest is defined as $$\widehat Q(\beta) = \frac{1}{n}\sum_{i=1}^n h(y_i,x_i'\beta)$$ where $h$ is convex and  $1$-Lipschitz with respect the second argument, $|h(y,t)-h(y,t')|\leq |t - t' |.$
We assume $h$ is differentiable in the second argument with derivative denoted $\nabla h$ to simplify exposition.  Let the true model parameter be defined by  $ \beta_0 \in \arg\min_{\beta\in\RR^p} \Ep[ h(y_i,x_i'\beta) ]$, and consequently we have $\Ep[ x_i \nabla h(y_i,x_i'\beta_0) ] = 0$.
The $\ell_1$-regularized estimator is given by the solution of
$$ \min_{\beta\in \RR^p} \widehat Q(\beta) + \frac{\lambda}{n} \|\beta\|_1. $$
 Under high level conditions \citen{vdGeer} derived bounds on the excess forecasting loss, $\Ep[h(y_i,x_i'\hat \beta)] - \Ep[h(y_i,x_i'\beta_0)]$,
 under sparsity-related assumptions,  and
 also specialized the results to logistic regression, density estimation, and other problems.\footnote{Results in other norms
 of interest could also be derived, and the behavior of the post-$\ell_1$-regularized estimators would also be interesting to consider.
 This is an interesting venue for future work.}  The choice of penalty parameter $\lambda$ derived in \citen{vdGeer} relies on using
 the contraction inequalities of \citen{LedouxTalagrandBook} in order to bound the score:
\begin{equation}\label{vdgeer} n\|\nabla \widehat Q(\beta_0)\|_\infty = \left\|\sum_{i=1}^nx_i\nabla h(y_i,x_i'\beta_0)\right\|_\infty \lesssim_P \left\|\sum_{i=1}^nx_i\xi_i\right\|_\infty,
\end{equation} where $\xi_i$ are independent Rademacher random variables, $P(\xi_i=1)=P(\xi_i=-1)=1/2$.
Then \citen{vdGeer} suggests further bounds on the right side of (\ref{vdgeer}). For efficiency reasons, we suggest simulating the $1-\gamma$ quantiles of  the right side of (\ref{vdgeer}) conditional on regressors. In either way one can achieve the domination of ``noise"
$\lambda/n \geq c\|\nabla \widehat Q(\beta_0)\|_\infty$ with high probability.  Note that since $h$ is 1-Lipschitz, this choice
of the penalty level is pivotal.

%
%

\section{Estimation Results for High Dimensional Sparse Models}\label{Sec:ResultsHDSM}

\subsection{Convergence Rates for Lasso and Post-Lasso}

Having introduced Condition ASM and the target parameter defined via (\ref{oracle}), our task becomes to estimate $\beta_0$.
We will focus on convergence results in the {\it prediction  norm} for $\delta = \widehat \beta - \beta_0$, which measures the accuracy of predicting $x_i'\beta_0$  over the design points $x_1,\ldots,x_n$,
$$
\|\delta\|_{2,n} := \sqrt{ \En[(x_i'\delta)^2] } = \sqrt{\delta '\En[ x_ix_i'] \delta }.$$

The prediction norm directly depends on the the Gram matrix $\En [x_ix_i']$.  Whenever $p>n$, the empirical Gram matrix $\En[x_ix_i']$ does not
have full rank and in principle is not well-behaved. However, we only need good behavior of certain moduli of continuity of the Gram matrix called  sparse eigenvalues.
We define the minimal $m$-sparse eigenvalue of a semi-definite matrix $M$ as
\begin{equation}\label{Def:RSE1}
\semin{m}[M] : = \min_{\|\delta \|_{0} \leqslant m, \delta \neq 0
 } \frac{  \delta 'M \delta  }{\|\delta\|^2},
\end{equation}
and the maximal $m$-sparse eigenvalue as
\begin{equation}\label{Def:RSE2}
 \semax{m}[M] : = \max_{\|\delta \|_{0} \leqslant m, \delta \neq 0
 } \frac{ \delta 'M \delta }{\|\delta\|^2},
\end{equation}
To assume that $\semin{m}[\En [x_ix_i']] >0$ requires that all empirical Gram submatrices formed by any $m$ components of $x_i$  are positive definite. To simplify asymptotic statements for Lasso and Post-Lasso, we use the following condition:

\textbf{Condition SE.} \textit{ There is $\ell_n \to \infty$ such that
$$\kappa' \leq \semin{\ell_n s}[\En [x_ix_i']] \leq \semax{\ell_n s}[\En [x_ix_i']] \leq \kappa'',$$
where $0< \kappa' <  \kappa'' < \infty$ are constants that do not depend on $n$.
}

\begin{remark}
It is well-known that Condition SE is quite plausible for many designs of interest. For instance, Condition SE holds with probability approaching one as $n \to \infty$ if $x_{i}$ is a normalized form of $\tilde x_i$, namely   $x_{ij}= \tilde x_{ij}/\sqrt{\En[\tilde x_{ij}^2]}$, and
\begin{itemize}
\item $\tilde x_i$, $i = 1,\ldots,n$, are i.i.d. zero-mean Gaussian random vectors that have population Gram matrix $\Ep[\tilde x_i \tilde x_i']$ with ones on the diagonal and its minimal and maximal $s\log n$-sparse eigenvalues bounded away from zero and from above, where $s\log n = o(n/\log p)$;
\item $\tilde x_i$, $i=1,\ldots,n$, are i.i.d. bounded zero-mean random vectors with $\| \tilde x_i\|_\infty \leq K_n$ a.s. that have population Gram  matrix $\Ep[\tilde x_i \tilde x_i']$ with ones on the diagonal and its minimal and maximal $s\log n$-sparse eigenvalues bounded from above and away from zero, where $K_n^2s\log^5(p\vee n)=o(n)$.
\end{itemize}
Recall that a standard assumption in econometric research is to assume that the population Gram matrix $\Ep[x_i x_i']$ has eigenvalues bounded from above and below, see e.g. \citen{newey:series}. The conditions above allow for this and more general behavior, requiring only that the $s \log n$ sparse eigenvalues of the population Gram matrix $\Ep[x_i x_i']$ are bounded from below and from above. The latter is important for allowing functions $x_i$ to be formed as a combination of elements from different bases, e.g. a combination of B-splines with polynomials.  \qed \end{remark}

The following theorem describes the rate of convergence for feasible Lasso in the Gaussian model under
Conditions ASM and SE. We formally  define the \emph{feasible Lasso} estimator $\hat\beta$ as either
the Iterated Lasso with penalty level given by
$X$-independent rule (\ref{Def:LambdaLASSOboound}) or $X$-dependent rule (\ref{Def:LambdaLASSO}) or Square-root Lasso  with penalty level given by  $X$-dependent rule (\ref{our penalty}) or $X$-independent  rule (\ref{our penalty asymptotic}), with the confidence level $1-\conflvl$ such that
\begin{equation}\label{Def: conf level}
\conflvl = o(1) \textrm{ and } \log(1/\conflvl) \lesssim \log (p \vee n).
\end{equation}

\begin{theorem}[Rates for Feasible Lasso]\label{corollary1:rate}
Suppose that conditions ASM and SE hold.  
Then for $n$ large enough the following bounds hold with probability at least $1-\gamma$:
 $$C' \|\widehat\beta - \beta_0\| \leq \|\widehat \beta -\beta_0 \|_{2,n} \leq C \sigma \sqrt{\frac{s\log (2p/\gamma)}{n}},$$
  where $C>0$ and $C'>0$ are constants, $C' \gtrsim \sqrt{\kappa'}$ and $C \lesssim 1/\sqrt{\kappa'}$,
 and $\log (p/\gamma)\lesssim \log (p \vee n)$.
\end{theorem}

\begin{remark} Thus the rate for estimating $\beta_0$ is $\sqrt{s/n}$, i.e. the root
of the number of parameters $s$ in the ``true" model divided by the sample size $n$, times a logarithmic
factor $\sqrt{ \log (p \vee n)}$. The latter factor can be thought of as the price of not knowing the ``true" model.   Note that the rate
for estimating the regression function $f$ over design points follows from the triangle inequality and Condition ASM:
\begin{equation}\label{Def:NORM_ER_OLD}
\sqrt{\En[ (f(z_i) - x_i'\hat \beta)^2]} \leqslant  \|\widehat \beta - \beta_0\|_{2,n}+c_s \lesssim_P  \sigma \sqrt{\frac{s\log (p \vee n)}{n}}.
\end{equation}
\end{remark}

\begin{remark} The result of Theorem  \ref{corollary1:rate} is an extension of
the results in the fundamental work of \citen{BickelRitovTsybakov2009} and \citen{MY2007} on infeasible Lasso
and \citen{CandesTao2007} on the Dantzig estimator. The result of Theorem  \ref{corollary1:rate} is derived
in \citen{BC-PostLASSO} for Iterated Lasso,  and in \citen{BCW-SqLASSO} and \citen{BCW-SqLASSO2}
for Square-root Lasso (with constants $C$ given explicitly).
Similar results also hold for  $\ell_1$-QR \cite{BC-SparseQR} and other M-estimation problems \cite{vdGeer}.
The bounds of Theorem \ref{corollary1:rate} allow the constructions of confidence sets for $\beta_0$, as  noted in \citen{Chern:SG}; see also
\citen{gautier:tsybakov}.  Such confidence sets rely on efficiently bounding $C$. Computing bounds for $C$
requires computation of combinatorial quantities depending
on the unknown model $T$ which makes the approach  difficult in practice.
In the subsequent sections,
we will present completely different approaches to inference which
have provable confidence properties for parameters of interest
and which are computationally tractable.  \qed
\end{remark}

As mentioned before, $\ell_1$-regularized estimators have an inherent bias towards zero and Post-Lasso was proposed to remove this
bias, at least in part. It turns out that we can bound the performance of Post-Lasso  as a function of Lasso's rate of convergence and Lasso's model selection ability.  For common designs, this bound implies that Post-Lasso performs at least as well as Lasso, and it can be strictly better in some cases.  Post-Lasso also has a smaller shrinkage bias than Lasso by construction.

The following theorem applies to any Post-Lasso estimator $\widetilde \beta$ computed using the model $\widehat T = \text{support}(\hat \beta)$
selected by a Feasible Lasso estimator $\widehat \beta$ defined before Theorem \ref{corollary1:rate}.

\begin{theorem}[Rates for Feasible Post-Lasso]\label{corollary3:postrate} Suppose the conditions of Theorem \ref{corollary1:rate} hold and let $\varepsilon>0$. Then there are constants $C'$ and $C_\varepsilon$ such that with probability $1-\gamma$
$$
\hat s = | \widehat T | \leq C' s,
$$
and with probability $1-\gamma-\varepsilon$
\begin{equation}
\sqrt{\kappa'} \|\widetilde \beta - \beta_0\| \leq \|\widetilde \beta - \beta_0\|_{2,n} \leq \ \ C_\varepsilon \sigma \sqrt{ \frac{ s \log (p \vee n)}{n} }.
\end{equation}
If further $|\|\widehat\beta\|_0-s|= o(s)$ and $ T \subseteq \widehat T$ with probability approaching one, then
\begin{equation}\label{superior to Lasso}
\|\widetilde \beta - \beta_0\|_{2,n} \lesssim_P \ \ \sigma \[\sqrt{ \frac{ o(s)  \log (p \vee n)}{n} } +  \sqrt{ \frac{ s }{n} } \].
 \end{equation}
If $\widehat T = T $ with probability approaching one, then Post-Lasso achieves the oracle performance
\begin{equation}\label{becomes oracle}
\|\widetilde \beta - \beta_0\|_{2,n} \lesssim_P \  \sigma \sqrt{ s/n }.
\end{equation}
\end{theorem}

\begin{remark} The theorem above shows that Feasible Post-Lasso achieves the same near-oracle rate as Feasible Lasso. Notably, this occurs despite the fact that Feasible Lasso may in general fail to correctly select the oracle model $T$ as a subset, that is $T \not \subseteq \widehat T$.  The intuition for this result is that any components of $T$ that Feasible Lasso misses are very unlikely to be important. Theorem \ref{corollary3:postrate} was derived in
\citen{BC-PostLASSO} and \citen{BCW-SqLASSO2}. Similar results have been shown before for
$\ell_1$-QR \cite{BC-SparseQR}, and can be derived for other methods that yield sparse estimators. \qed
\end{remark}

\subsection{Monte Carlo Example}

In this section we compare the performance of various estimators relative to the ideal oracle linear regression estimator.
The oracle estimator applies ordinary least square to the true model by regressing the outcome on only the control variables with non-zero coefficients. Of course, the oracle estimator is not available outside Monte Carlo experiments.

We considered the following regression model:
$$ y = x'\beta_0 + \epsilon, \  \ \beta_0 =(1,1,1/2,1/3,1/4,1/5,0,\ldots,0)', $$
where $x = (1,z')'$ consists of an intercept and covariates $z \sim N(0,\Sigma)$,
and the errors $\epsilon$ are independently and identically
distributed $\epsilon \sim N(0,\sigma^2)$. The dimension $p$ of the covariates $x$ is $500$, and the dimension $s$ of the true model is $6$. The sample size $n$ is $100$.   The regressors are correlated with $\Sigma_{ij} = \rho^{|i-j|}$ and $\rho = .5$. We consider the levels of noise to be $\sigma = 1$ and $\sigma=0.1$. For each repetition we draw new $x$'s and $\epsilon$'s.

We consider infeasible Lasso and Post-Lasso estimators, feasible Lasso and Post-Lasso estimators described in the previous
section, all with X-dependent penalty levels, as well as (5-fold) cross-validated (CV) Lasso and Post-Lasso. We summarize results on estimation performance in Table \ref{Table:MC} which records for each estimator $\bar \beta$  the norm of the bias $\|\Ep[ \bar \beta - \beta_0]\|$ and also the empirical risk $\{\Ep[(x_i'( \bar \beta - \beta_0 ))^2]\}^{1/2}$ for recovering the regression function. In this design, infeasible Lasso, Square-root Lasso, and Iterated Lasso exhibit substantial bias toward zero. This bias is somewhat alleviated by choosing the penalty-level via cross-validation, though the remaining bias is still substantial. It is also apparent that, as intuition and theory would suggest, the post-penalized estimators remove a large portion of this shrinkage bias.
We see that among the feasible estimators,  the best performing methods are the Post-Square-root Lasso and Post-Iterated Lasso. Interestingly,
cross-validation also produces a Post-Lasso estimator that performs nearly as well, although the procedure is much more expensive computationally.
The Post-Lasso estimators perform better than Lasso estimators primarily due to a much lower shrinkage bias which is beneficial in the design considered.

{\small\begin{table}[ht]\begin{tabular}{lccccc}
\hline
& \multicolumn{2}{c}{{\bf High Noise ($\sigma = 1$)}} & & \multicolumn{2}{c}{{\bf Low Noise ($\sigma = 0.1$)}} \\
Estimator &   Bias  & Prediction Error & & Bias & Prediction Error\\ \hline
Lasso  &  0.444 & 0.654 & &  0.0487 & 0.0700\\
Post-Lasso  &  0.129 & 0.347 & &  0.0054  & 0.0300\\
Square-root Lasso  &  0.526 & 0.770& &  0.0615  &  0.0870  \\
Post-Square-root Lasso  &  0.187 & 0.364 & &  0.0035  & 0.0238  \\
Iterated Lasso & 0.437 & 0.644 & & 0.0477 &  0.0687 \\
Post-Iterated Lasso & 0.133 & 0.360 & & 0.0056  & 0.0297  \\
CV Lasso & 0.265 & 0.516 & & 0.0233 & 0.0987\\
CV Post-Lasso & 0.148 & 0.415 & & 0.0035 & 0.0237\\
Oracle   &  0.035  & 0.238 & &  0.0035  & 0.0237 \\
\hline
\\
\end{tabular}
\caption{\footnotesize The table displays the mean bias and the mean prediction error.  The average number of components selected by Lasso was $5.18$ in the high noise case and $6.44$ in the low noise case. In the case of CV Lasso, the average size of the model was $29.6$ in the high noise case and $10.0$ in the low noise case. Finally, the CV Post-Lasso selected models with average size of $7.1$ in the high noise case and $6.0$ in the low noise case.} \label{Table:MC}
\end{table}}

\section{Inference on Structural Effects with High-Dimensional Instruments}\label{Sec:IV}

\subsection{Methods and Theoretical Results}
In this section, we consider the linear instrumental variable (IV) model with many instruments.
Consider the Gaussian simultaneous equation model:
\begin{eqnarray}
& &  y_{1i}  = y_{2i} \alpha_{1} + w_i'\alpha_2 + \zeta_i, \label{Def: second stage} \\
& & y_{2i}  = f(z_i) + v_i,  \label{Def: first stage} \\
&  &  \(\begin{array}{cc} \zeta_i \\ v_i\end{array}\)  \mid z_i  \sim N\(0,\(\begin{array}{cc} \sigma^2_\zeta & \sigma_{\zeta v} \\ \sigma_{\zeta v} & \sigma^2_{v}\end{array}\)\). \label{Def: IVerrors}
\end{eqnarray}
Here $y_{1i}$ is the response variable, $y_{2i}$ is the endogenous variable,
$w_i$ is a $k_w$-vector of control variables,
 $z_i = (u_i',w_i')'$ is a vector of instrumental variables (IV), and $(\zeta_i, v_i)$ are disturbances that
are independent of $z_i$.  The function $f(z_i) = \Ep[y_{2i}|z_i]$, the optimal instrument, is an unknown, potentially complicated function of the elementary instruments $z_i$.  The main parameter of interest is the coefficient on $y_{2i}$, whose true value is $\alpha_1$.
 We treat $\{z_i\}$ as fixed throughout.

Based on these elementary instruments, we create a high-dimensional vector of technical instruments, $x_i = P(z_i)$, with dimension
$p$ possibly much larger than the sample size though restricted via conditions stated below. We then estimate the
the optimal instrument $ f(z_i)$ by
 \begin{equation}\label{Def: IV-Lasso1}
\widehat f(z_i) = x_i'\widehat \beta,
 \end{equation}
where  $\hat \beta$ is a feasible Lasso or Post-Lasso estimator as formally defined in
the previous section.

Sparse-methods take advantage of approximate sparsity and ensure that many elements of $\widehat\beta$ are zero when $p$ is large. In other words, sparse-methods will select a small subset of the available technical instruments.  Let  $A_i = (f(z_i), w_i')'$ be the ideal instrument vector, and let
 \begin{equation}\label{Def: IV-Lasso2}
\widehat A_i =  ( \widehat f(z_i), w_i')'
 \end{equation}
be the estimated instrument vector.  Denoting $d_i=(y_{2i},w_i')'$, we form the feasible IV estimator using the estimated instrument vector as
 \begin{equation}\label{Def: IV-Lasso3}
\widehat \alpha^* =  \Big(\En [\widehat A_i d_i']  \Big)^{-1}  \Big( \En [\widehat A_i y_{1i}]    \Big).
 \end{equation}

The main regularity condition is recorded as follows.

\textbf{Condition ASIV.} \emph{In the linear IV model
(\ref{Def: second stage})-(\ref{Def: IVerrors}) with technical instruments $x_i = P(z_i)$, the following assumptions hold:
 (i) the parameter values $\sigma_v$, $\sigma_\zeta$ and the eigenvalues of $Q_n=\En[A_iA_i']$ are bounded away from zero and from above uniformly in $n$, (ii) condition ASM holds for (\ref{Def: first stage}),  namely for each $i =1,...,n$, there exists $\beta_0 \in \Bbb{R}^p$, such that
 $
 f(z_i) = x_i'\beta_0+ r_i,   \ \  \|\beta_0\| \leq s,  \ \ \{ \En[r_i^2] \}^{1/2} \leq K \sigma_{v} \sqrt{s/n},
 $
where constant $K$ does not depend on $n$,
(iii) condition SE holds for $\En[x_ix_i']$,  and (iv)  $s^2\log^2 (p\vee n) = o(n)$.}

The main inference result is as follows.

\begin{theorem}[Asymptotic Normality for IV Estimator Based on Lasso and Post-Lasso]\label{Thm:IV} Suppose Condition ASIV holds.
The IV estimator constructed in (\ref{Def: IV-Lasso3}) is $\sqrt{n}$-consistent and is asymptotically efficient, namely as $n$ grows:
$$
(\sigma^2_\zeta Q_n^{-1})^{-1/2} \sqrt{n}(\widehat \alpha^* - \alpha) = N(0, I) + o_P(1),
$$
and the result also holds with $Q_n$ replaced by $\widehat Q_n= \En [\widehat A_i \widehat A_i']$
 and $\sigma^2_\zeta$ by $\hat \sigma^2_\zeta = \En[ (y_{1i} - \widehat A_i'\widehat \alpha^*)^2]$.
\end{theorem}

\begin{remark} The theorem shows that the IV estimator based on estimating the first-stage with Lasso or Post-Lasso is asymptotically as efficient as the infeasible optimal IV estimator that uses $A_i$ and thus achieves the  semi-parametric efficiency bound of \citen{chamberlain}.
\citen{BellChernHans:Gauss} show that the result continues to hold  when other sparse methods are used
 to estimate the optimal instruments.
The sufficient conditions for showing the IV estimator obtained using sparse-methods to estimate the optimal instruments is asymptotically efficient include a set of technical conditions and the following key growth condition:
$ s^2 \log^2 (p\vee n) = o(n).$
This rate condition requires the optimal
instruments to be sufficiently smooth  so that a relatively small number of series terms can be used
to approximate them well.  This smoothness ensures that the impact of instrument estimation on the IV estimator is asymptotically negligible. The rate condition $s^2\log^2 (p\vee n) = o(n)$ can be substantive and cannot be substantially weakened for the full-sample IV estimator considered above.  However, we can replace this condition with the weaker condition that
$ s \log (p\vee n) = o(n)$
by employing a sample splitting method from the many instruments literature \cite{AngristKruegerSplitSample1995} as established in \citen{BellChernHans:Gauss} and \citen{BellChenChernHans:nonGauss}.
 Moreover,  \citen{BellChenChernHans:nonGauss} show that the result
 of the theorem, with some appropriate modifications, continues to apply under heteroscedasticity though the estimator does not necessarily attain the semi-parametric efficiency bound.
 In order to achieve full efficiency allowing for heteroscedasticity, we would need to estimate the conditional variance of the
 structural disturbances in the second stage equation. In principle, this estimation could be done using sparse methods.
 $\qed$ \end{remark}

\subsection{Weak Identification Robust Inference with Very Many Instruments}
Consider the simultaneous equation model:
\begin{eqnarray}
& &  y_{1i}  = y_{2i} \alpha_{1} + w_i'\alpha_2 + \zeta_i, \ \ \zeta_i \mid z_i   \sim N\(0, \sigma^2_\zeta\), \label{Def: WIVmodel}
\end{eqnarray}
where $y_{1i}$ is the response variable, $y_{2i}$ is the endogenous variable,  $w_i$ is a $k_w$-vector of control variables, $z_i = (u_i',w_i')'$ is a vector of instrumental variables (IV), and $\zeta_i$ is a disturbance that is independent of $z_i$. We treat $\{z_i\}$ as fixed throughout.

We  would like to  use a high-dimensional vector $x_i=P(z_i)$ of technical instruments for inference that is robust to weak identification.  We propose a method for inference based on inverting pointwise tests
performed using a sup-score statistic defined below.  The procedure is similar in spirit to \citen{anderson:rubin} and \citen{ss:weakiv} but uses a very different statistics that is well-suited to cases with very many instruments.

In order to formulate the sup-score
statistic, we first partial-out the effect of controls $w_i$ on the key variables.
For an $n$-vector $\{u_i, i=1,...,n\}$, define $\tilde u_i = u_i - w_i'\En[w_i w_i']^{-1} \En[w_i u_i]$, i.e. the residuals left after regressing this vector on $\{w_i, i=1,...,n\}$. Hence $\tilde y_{1i}$, $\tilde y_{2i}$, and $\tilde x_{ij}$ are residuals
obtained by partialling out controls.  Also, let $\tilde x_i = (\tilde x_{i1},...,\tilde x_{ip})'$. In this formulation, we
omit elements of $w_i$ from $\tilde x_{ij}$ since they are eliminated by partialling out. We then normalize without loss of generality
\begin{equation}\label{def: normalize tilde X}
\En[\tilde x_{ij}^2] =1, \ \ j =1,...,p.
\end{equation}
  The sup-score statistic for testing
the hypothesis $\alpha_1 = a $ takes the form:
$$
\Lambda_a =  \max_{1 \leq j \leq p} \frac{|n \En [(\tilde y_{1i} - \tilde y_{2i}a) \tilde x_{ij}]|}{\sqrt{\En[
(\tilde y_{1i} - \tilde y_{2i}a)^2 \tilde x^2_{ij} ]}}.
$$
If the hypothesis $\alpha_1 = a$ is true, then the critical value for achieving level $\gamma$ is
\begin{equation}
\Lambda(1- \gamma|W,X) =  1- \gamma-\text{\rm quantile  of  } \ \ \max_{1 \leq j \leq p} \frac{|n \En [ \tilde g_i \tilde x_{ij}]|}{\sqrt{\En[ \tilde g^2_i \tilde x^2_{ij} ]}} \mid W, X
 \end{equation}
 where $W=[w_1,...,w_n]'$, $X=[x_1,...,x_n]'$, and $g_1,...,g_n$ are i.i.d. $N(0,1)$ variables
 independent of $W$ and $X$; $\tilde{g}_i$ denotes the residuals left after projecting $\{g_i\}$ on $\{w_i\}$ as defined above.  We can approximate the critical value
 $\Lambda(1- \gamma|W,X)$ by simulation conditional on $X$ and $W$.  It is also possible to use a simple asymptotic bound on this critical value of the form
\begin{equation}
\Lambda(1- \gamma):= c \sqrt{n}\Phi^{-1}(1- \gamma/2p)  \le c  \sqrt{2 n \log(2p/\gamma)},
 \end{equation}
 for $c > 1$. The finite-sample $(1- \gamma)$ -- confidence region for $\alpha_1$ is then given by
$$
\mathcal{C}:=\{ a \in \Bbb{R}:  \Lambda_a \leq \Lambda(1- \gamma|W,X)\},
$$
while a large sample  $(1- \gamma)$ -- confidence region is given by
$
\mathcal{C}':= \{ a \in \Bbb{R}:  \Lambda_a \leq \Lambda(1- \gamma)\}.
$

The main regularity condition is recorded as follows.

\textbf{Condition HDIV.} \emph{Suppose the linear IV model
(\ref{Def: WIVmodel}) holds.
Consider the $p$-vector of instruments $x_i = P(z_i)$, $i=1,...,n$,
such that $(\log p)/n \to 0$. Suppose further that the following assumptions hold uniformly in $n$:
 (i) the parameter value $\sigma_\zeta$  is bounded away from zero and from above,
(ii) the dimension of $w_i$ is bounded and the eigenvalues
of the Gram matrix $\En[w_i w_i']$ are bounded away from zero,
(iii) $\|w_i\| \leq K$ and $|\tilde x_{ij}| \leq K$ for all $1\leq i \leq n$ and all $1 \leq j \leq p$,
where $K$ is a constant, independent of $n$.
}

The main inference result is as follows.

\begin{theorem}[Valid Inference based on the Sup-Score Statistic]\label{Thm:WIV}
(1) Suppose the linear IV model (\ref{Def: WIVmodel}) holds.  Then
$ \Pr( \alpha_1 \in \mathcal{C} ) = 1- \gamma$. (2)
Suppose further that condition HDIV holds, then
$ \Pr( \alpha_1 \in \mathcal{C}' ) \geq 1- \gamma -o(1)$.  (3) Moreover, if $a$ is such that
that
$$
\max_{1 \leq j \leq p} \frac{|a- \alpha_1|  \sqrt{n} |\En[ \tilde y_{2i} \tilde x_{ij} ]|/\sqrt{ \log p}}{\sigma_{\zeta} + |a- \alpha_1|\sqrt{\En[\tilde y_{2i}^2 \tilde x_{ij}^2]}}
 \to \infty,
$$
 then
$ \Pr ( a \in \mathcal{C}) = o(1)$ and $\Pr (a \in \mathcal{C}') = o(1)$.
\end{theorem}

\begin{remark} The theorem shows that the confidence regions $\mathcal{C}$ and $\mathcal{C}'$
constructed above have finite-sample and large sample validity, respectively.
Moreover, the probability of including a false point $a$ in either $\mathcal{C}$ or $\mathcal{C}'$
 tends to zero as long as $a$ is sufficiently distant from $\alpha_1$ and instruments
are not too weak.  In particular, if there is a strong instrument, the confidence regions will eventually
exclude points $a$ that are further than $\sqrt{ (\log p)/n}$ away from $\alpha_1$. Moreover,
if there are instruments whose correlation with the endogenous variable is of greater order
than $\sqrt{ (\log p)/n}$, then the confidence regions will asymptotically be bounded.  Finally,
note that a nice feature of the construction is that it provides provably valid
confidence regions and does not require computation of some combinatorial quantities, in sharp
contrast to other recent proposals for inference, e.g. \citen{gautier:tsybakov}.  Lastly, we note that it is not difficult
to generalize the results to allow for an increasing number of controls $w_i$ under suitable
technical conditions that restrict the number of controls and their envelope in relation to the sample size. Here we did
not consider this possibility in order to highlight the impact of very many instruments
more clearly.  The result (2) extends to non-Gaussian, heteroscedastic cases;
we refer to \citen{BellChenChernHans:nonGauss} for relevant details.   \qed
\end{remark}

\begin{remark}[Inverse Lasso Interpretation] The construction of confidence regions above can be given the following \emph{Inverse Lasso} interpretation.
Let
$$
\hat \beta_a  = \arg\min_{\beta \in \Bbb{R}^p} \En[ (\tilde y_{1i} - a \tilde y_{2i}) - \tilde x_{ij}'\beta]^2 + \frac{\lambda}{n}
\sum_{j=1}^p | \beta_j | \gamma_{aj} , \ \ \gamma_{aj} = \sqrt{\En[ (\tilde y_{1i} - \tilde y_{2i}a)^2 \tilde x^2_{ij} ]}.
$$
If $\lambda = 2\Lambda(1- \gamma|W,X)$, then $ \mathcal{C}$ is equivalent to the region $\{ a \in \Bbb{R}:  \hat \beta_a = 0\}$.
If $\lambda = 2\Lambda(1- \gamma)$, then $ \mathcal{C}'$ is equivalent to the region $\{ a \in \Bbb{R}:  \hat \beta_a = 0\}$.  In words,
to construct these confidence regions, we collect all potential values of the structural parameter, where the Lasso regression of the potential structural disturbance on the instruments yields zero coefficients on the instruments.  This idea is akin to the Inverse Quantile Regression
and Inverse Least Squares ideas in \citen{ch:iqrWeakId} and \citen{ch:WeakId}. \qed
\end{remark}

\subsection{Monte Carlo Example: Instrumental Variable Model}

The theoretical results presented in the previous sections suggest that using Lasso to aid in fitting the first-stage regression should result in IV estimators with good estimation and inference properties.  In this section, we provide simulation evidence on these properties of IV estimators using iterated Lasso to select instrumental variables for a second-stage estimator.  We also considered Square-root Lasso for variable selection.  The results were similar to those for iterated Lasso, so we report only the iterated Lasso results.

Our simulations are based on a simple instrumental variables model of the form
$$\begin{array}{ll}
y_{1i} &= \alpha y_{2i} + \zeta_i \\
y_{2i} &= x_i'\Pi + v_i \\
\end{array}  \ \ \ \left(\begin{array}{ll}\zeta_i \\ v_i \end{array}\right)\mid x_i \sim N\(0,\(\begin{array}{cc} \sigma^2_{\zeta} & \sigma_{\zeta v} \\ \sigma_{\zeta v} & \sigma^2_{v}\end{array}\)\) \ {\rm i.i.d.,}
$$
where $\alpha=1$ is the parameter of interest, and $x_i = (x_{i1},...,x_{i100})' \sim N(0,\Sigma_X)$ is the instrument vector with $E[x_{ih}^2] = \sigma^2_x$ and $\mbox{Corr}(x_{ih},x_{ij}) = .5^{|j-h|}$.  In all simulations, we set $\sigma^2_{\zeta} = 1$ and $\sigma^2_x = 1$.  We also use $\mbox{Corr}(\zeta,v$) = .3.

We consider several different settings for the other parameters.  We provide simulation results for sample sizes, $n$, of 100 and 500.  In one simulation design, we set $\Pi = 0$ and $\sigma^2_v = 1$.  In this case, the instruments have no information about the endogenous variable, so $\alpha$ is unidentified.  We refer to this as the ``No Signal'' design.  In the remaining cases, we use an ``exponential'' design for the first stage coefficients, $\Pi$, that sets the coefficient on $x_{ih} = .7^{h-1}$ for $h=1,...,100$ to provide an example of Lasso's performance in settings where the instruments are informative.  This model is approximately sparse, since the majority of explanatory power is contained in the first few instruments, and obeys the regularity conditions put forward above.  We consider values of $\sigma^2_v$ which are chosen to benchmark three different strengths of instruments.  The three values of $\sigma^2_v$ are found as $\sigma^2_v = \frac{n \Pi'\Sigma_Z\Pi}{F^*\Pi'\Pi}$ for $F^*$ of 10, 40, or 160.

For each setting of the simulation parameter values, we report results from several estimation procedures.  A simple possibility when presented with $p < n$ instrumental variables is to just estimate the model using 2SLS and all of the available instruments.  It is well-known that this will result in poor-finite sample properties unless there are many more observations than instruments; see, for example, \citen{bekker}.  Fuller's \citeyear{fuller} estimator (FULL)\footnote{The Fuller estimator requires a user-specified parameter.  We set this parameter equal to one which produces a higher-order unbiased estimator.  See \citen{hhk:weakmse} for additional discussion.} is robust to many instruments as long as the presence of many instruments is accounted for when constructing standard errors and $p < n$; see \citen{bekker} and \citen{hhn:weakiv} for example.  We report results for these estimators in rows labeled 2SLS(All) and FULL(All) respectively.\footnote{All models include an intercept.  With $n = 100$, we randomly select 98 instruments to use for 2SLS(All) and FULL(All).}   In addition, we report Fuller and IV estimates based on the set of instruments selected by Lasso with two different penalty selection methods.  IV-Lasso and FULL-Lasso are respectively 2SLS and Fuller using instruments selected by Lasso with penalty obtained using the iterated method outlined in Appendix A.  We use an initial estimate of the noise level obtained using the regression of $y_2$ on the instrument that has the highest simple correlation with $y_2$. IV-Lasso-CV and FULL-Lasso-CV are respectively 2SLS and Fuller using instruments selected by Lasso using 10-fold cross-validation to choose the penalty level.  We also report inference results based on the Sup-Score test developed in Section 5.2.

In Table \ref{Table:IVmc}, we report root-mean-squared-error (RMSE), median bias (Med. Bias), rejection frequencies for 5\% level tests (rp(.05)), and the number of times the Lasso-based procedures select no instruments ($\|\widehat\Pi\|_0 = 0$).  For computing rejection frequencies, we estimate conventional 2SLS standard errors for all 2SLS estimators, and the many instrument robust standard errors of \citen{hhn:weakiv} for the Fuller estimators.  In cases where Lasso selects no instruments, the reported Lasso point estimation properties are based on the feasible procedure that enforces identification by lowering the penalty until one variable is selected.  Rejection frequencies in cases where no instruments are selected are based on the feasible procedure that uses conventional IV inference using the selected instruments when this set is non-empty and otherwise uses the Sup-Score test.

The simulation results show that Lasso-based IV estimators is useful in situations with many instruments.
As expected, 2SLS(All) does extremely poorly along all dimensions.  FULL(All) also performs worse than the Lasso-based estimators in terms of estimator risk (RMSE) in all cases.  The Lasso-based procedures do not dominate FULL(All) in terms of median bias, though all of the Lasso-based procedures have smaller median bias than FULL(All) when $n = 100$ and there is some signal in the instruments and are very similar with $n = 500$. In terms of size of 5\% level tests, we see that the Sup-Score test uniformly controls size as indicated by the theory.  IV-Lasso and FULL-Lasso using the iterated penalty selection method also do a very good job controlling size across all of the simulation settings with a worst-case rejection frequency of .064 (with simulation standard error of .01) and the majority of rejection frequencies below .05.  Interestingly, when there is no signal in the instrument, the Lasso-based estimators using penalty selected by CV have substantial size-distortions when $n = 100$ which is due to the CV penalty being small enough that instruments are still selected despite there being no signal.  The iterated penalty is such that, at least approximately, only instruments whose coefficients are outside of a $\sqrt{n}$ neighborhood of 0 are selected and thus overselection in cases with little signal is guarded against.  Despite the problem with using CV when there is no signal, it is worth noting that the Lasso-based procedures with CV penalty produce tests with approximately correct size in all other parameter settings.

{\footnotesize \begin{table}[ht!]\title{\textbf{Instrumental Variables Model Simulation Results}}
\begin{tabular}{l|cccc|cccc}
\multicolumn{1}{c}{}          & \multicolumn{4}{c}{$n=100$} & \multicolumn{4}{c}{$n=500$}\\
Estimator &	RMSE &	Med. Bias &	 rp(.05)	& $\|\widehat{\Pi}\|_0 = 0$ &	 RMSE  &	 Med. Bias & rp(.05) & $\|\widehat{\Pi}\|_0 = 0$ \\
\hline
\multicolumn{9}{c}{No Signal}\\
\hline
2SLS(All)&0.318&0.305&0.862&   &0.312&0.297&0.852&\\
FULL(All)&2.398&0.248&0.704&   &1.236&0.318&0.066&\\
IV-Lasso &0.511&0.338&0.014&455 &0.477&0.296&0.012&486\\
FULL-Lasso&0.509&0.338&0.010&455 &0.477&0.296&0.012&486\\
IV-Lasso-CV&0.329&0.301&0.652&0  &0.478&0.299&0.064&348\\
FULL-Lasso-CV&0.359&0.305&0.384&0  &0.474&0.299&0.054&348\\
Sup-Score& & &0.004&                             & & &0.010 &\\
\hline
\multicolumn{9}{c}{$F^*=10$}\\
\hline
2SLS(All)&0.058&0.058&0.806&   &0.026&0.025&0.808&\\
FULL(All)&0.545&0.050&0.690&   &0.816&0.006&0.052&\\
IV-Lasso &0.055&0.020&0.042&147 &0.027&0.009&0.056&160\\
FULL-Lasso&0.054&0.020&0.032&147 &0.027&0.009&0.044&160\\
IV-Lasso-CV&0.052&0.024&0.072&10  &0.027&0.009&0.054&202\\
FULL-Lasso-CV&0.051&0.022&0.068&10  &0.027&0.009&0.044&202\\
Sup-Score& & &0.006&                              & & &0.004 &\\
\hline
\multicolumn{9}{c}{$F^*=40$}\\
\hline
2SLS(All)&0.081&0.072&0.626&   &0.036&0.032&0.636&\\
FULL(All)&0.951&0.050&0.690&   &0.038&0.000&0.036&\\
IV-Lasso &0.051&0.012&0.048&1 &0.022&0.003&0.048&0\\
FULL-Lasso&0.051&0.011&0.046&1 &0.022&0.002&0.038&0\\
IV-Lasso-CV&0.048&0.016&0.058&0  &0.022&0.004&0.052&0\\
FULL-Lasso-CV&0.049&0.014&0.050&0  &0.022&0.003&0.042&0\\
Sup-Score& & &0.004&                              & & &0.006 &\\
\hline
\multicolumn{9}{c}{$F^*=160$}\\
\hline
2SLS(All)&0.075&0.062&0.306&   &0.034&0.029&0.334&\\
FULL(All)&1.106&0.023&0.622&   &0.026&0.002&0.044&\\
IV-Lasso &0.049&0.005&0.064&0 &0.022&0.002&0.044&0\\
FULL-Lasso&0.049&0.002&0.056&0 &0.022&0.001&0.040&0\\
IV-Lasso-CV&0.048&0.006&0.054&0  &0.022&0.002&0.040&0\\
FULL-Lasso-CV&0.049&0.003&0.048&0  &0.022&0.000&0.038&0\\
Sup-Score& & &0.004&                              & & &0.010 &\\
\hline
\multicolumn{1}{c}{}\\
\end{tabular}\caption{{\footnotesize Results are based on 500 simulation replications. $F^*$ measures the strength of the instruments as outlined in the text. We report root-mean-square-error (RMSE), median bias (Med. Bias), rejection frequency for 5\% level tests (rp(.05)), and the number of times the Lasso-based procedures select no instruments ($\|\widehat\Pi\|_0 = 0$). Further details are provided in the text.
  } }\label{Table:IVmc}
\end{table}}

\begin{figure}[h!]
\centering
\includegraphics[width=\textwidth]{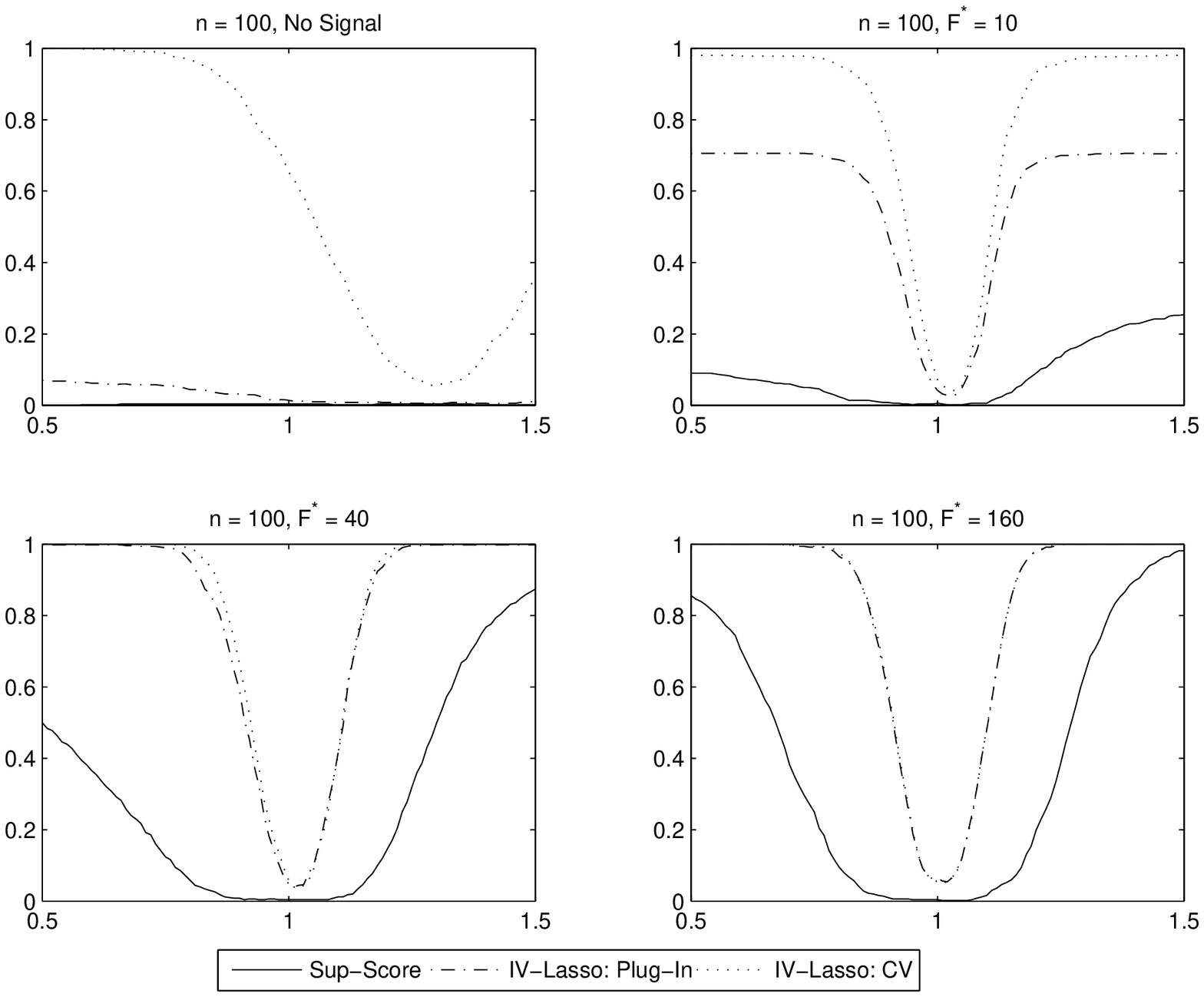}
\caption{Power curves for Sup-Score test, IV-Lasso with Iterated penalty, and IV-Lasso with penalty selected by 10-Fold Cross-Validation from IV simulation with 100 observations.}\label{Fig:IVSimPowerN100}
\end{figure}

To further examine the properties of the inference procedures that appear to give small size distortions, we plot the power curves of 5\% level tests using the Sup-Score test and IV-Lasso with the iterated and CV penalty choices with $n = 100$ in Figure \ref{Fig:IVSimPowerN100}.\footnote{The power curves in the $n = 500$ case are qualitatively similar.}  
We see that both the Sup-Score test and IV-Lasso using the iterated procedure augmented with Sup-Score test when no instruments are selected appear to uniformly control size and have some power against alternatives when the model is identified.  It is also clear that of these two procedures, the  IV-Lasso has substantially more power than the Sup-Score test.  The figures also show that IV-Lasso with iterated penalty has almost as much power as IV-Lasso using the CV penalty while avoiding the substantial size distortion and spurious power produced by using CV when there is no signal.  

Overall, the simulation results are favorable to the Lasso-based IV methods.  The Lasso-based estimators dominate the other estimators considered based on RMSE and have relatively small finite sample biases.  The Lasso-based procedures also do a good job in producing tests with size close to the nominal level.  There is some evidence that the Fuller-Lasso may do better than 2SLS-Lasso in terms of testing performance though these procedures are very similar in the designs considered.  It also seems that tests based on IV-Lasso using the iterated penalty selection rule may perform better than tests based on IV-Lasso using cross-validation to choose the Lasso penalty levels, especially when there is little explanatory power in the instruments.




\section{Inference on Treatment and Structural Effects Conditional on Observables}\label{Sec:Treatment}

\subsection{Methods and Theoretical Results}
We consider the following partially linear model,
\begin{eqnarray}\label{eq: PL1}
&  & y_{1i}  = d_i\alpha_0 + g(z_i) + \zeta_i, \\
& & d_i  = m(z_i) + v_i, \label{eq: PL2} \\
&  &  \(\begin{array}{cc} \zeta_i \\ v_i\end{array}\) \mid z_i   \sim N\(0,\(\begin{array}{cc} \sigma^2_\zeta & 0 \\ 0& \sigma^2_{v}\end{array}\)\) \label{Def: errors}
\end{eqnarray}
where $d_i$ is a policy/treament variable whose impact we would like to infer, and $z_i$ represents confounding factors on which we need to condition. This model is of interest in our international growth example discussed in the next section as well as in many empirical studies \cite{heckman:metricslabormarkets,imbens:review}.
The confounding factors affect the policy variable via $m(z_i)$.  We assume that $m(z_i)$ and $g(z_i)$ each admit an approximately sparse form and use linear combinations of technical control terms $x_i = P(z_i)$ to approximate them.

There are at least three obvious strategies for inference:
\begin{itemize}
\item[(i)] Estimate $\alpha_0$ by applying a Feasible Lasso method to model (\ref{eq: PL1}) without penalizing $\alpha_0$,
\item[(ii)] Estimate $\alpha_0$ by applying a Post-Lasso method to model (\ref{eq: PL1})  without penalizing $\alpha_0$,
\item[(iii)] Estimate $\alpha_0$ by applying an Indirect Post-Lasso where
$\alpha_0$ is estimated by running standard least squares regression of $y$ on $d$ and control terms selected in a preliminary Feasible Lasso regression of $d_i$ on $x_i$ in (\ref{eq: PL2}).
\end{itemize}
Note that it is most natural not to penalize $\alpha_0$ since the goal is to quantify the impact of $d_i$.  (The previous rate results derived in Theorems \ref{corollary1:rate} and \ref{corollary3:postrate} for the regression function extend to the case where the coefficients on a fixed number of variables are not penalized.) In what follows, we shall refer to options (i), (ii), and (iii) respectively as Lasso, Post-Lasso, and Indirect Post-Lasso.

Regarding inference, ``intuition" suggests that if $g$ can be estimated at faster than the $n^{1/4}$ rate then any of (i)-(iii) could be $\sqrt{n}$-consistent and asymptotically normal. It turns out that this ``intuition" is often correct for options (ii) and (iii) but is wrong for option (i). Indeed, it is possible to show that under rather strong regularity conditions that
\begin{equation}\label{Def: nonrobust}
(\sigma_{\zeta}^2 [\mathbb{E}_n v_i^2]^{-1})^{-1/2} \sqrt{n}(\bar \alpha - \alpha_0) =  N(0, 1 ) + o_{P}(1),
\end{equation}
where $\sigma_{\zeta}^2 [\mathbb{E}_n v_i^2]^{-1}$
is the semi-parametric efficiency bound for estimating $\alpha_0$, for $\bar \alpha$ denoting the estimators (ii) or (iii) above.  Unfortunately, the
distributional result (\ref{Def: nonrobust}) is not very robust to modest violations of regularity conditions and may provide a poor approximation to the finite-sample distributions of the estimators for $\alpha_0$.  The reason is that Lasso applied to
(\ref{eq: PL1}) may miss important terms relating $d_i$ to $z_i$ through $m(z_i)$ and thus suffer from substantial omitted variables bias.  On the other hand, Lasso applied only to (\ref{eq: PL2}), even if successful in selecting adequate controls for the relationship between $d_i$ and $z_i$, may miss important terms in $g(z_i)$ and thus be highly inefficient.  We illustrate this lack of robustness through a simulation experiment reported below.

Instead of using Lasso, Post-Lasso, or Indirect Post-Lasso, we advocate a  ``double-Post-Lasso'' method.  To define this estimator, we write the reduced
form corresponding to (\ref{eq: PL1})-(\ref{eq: PL2}):
\begin{eqnarray}
& & y_{1i}  = \alpha_0 m(z_i)+ g(z_i) +  \alpha_0v_i +\zeta_i,\label{eq: RPL2} \\
& & d_i  = m(z_i) + v_i. \label{eq: RPL2}
\end{eqnarray}
Now we have two equations and hence can apply Lasso methods to each equation to select control terms. That is,
we run  Lasso regression of $y_{1i}$ on $x_i=P(z_i)$ and Lasso regression of $d_i$ on $x_i=P(z_i)$.  Then we can run least squares of $y_{1i}$ on $d_i$ and
the union of the controls selected in each equation to estimate and perform inference on $\alpha_0$.
By using this procedure we  increase the chances for successfully recovering terms
that approximate the key control term $m(z_i)$, which results
in improved robustness properties. Indeed, the resulting procedure is
considerably more robust in computational experiments and requires much
weaker regularity conditions than the obvious strategies outlined above.

Now we formally define the double-Post-Lasso estimator.   Let $\hat I_1 = {\rm support }(\hat \beta_1)$
denote the control terms selected by a feasible Lasso estimator $\hat \beta_1$
computed using data $(y_i,x_i) = (d_{i}, x_i), i =1,...,n$.
Let $\hat I_2 = {\rm support }(\hat \beta_2)$ denote the control terms selected by a feasible Lasso estimator
$\hat \beta_2$ computed using  data $(y_i,x_i) = (y_{1i}, x_i), i =1,...,n$.
The double-Post-Lasso estimator $\check \alpha$ of $\alpha_0$ is defined as the least
squares  estimator obtained by regressing $y_{1i}$ on $d_i$ and the selected control terms
$x_{ij}$ with $j \in \hat I \supseteq \hat I_1 \cup \hat I_2$:
$$
(\check \alpha, \check \beta) = \underset{ \alpha \in \Bbb{R}, \beta \in \Bbb{R}^p}{\rm argmin}\{ \En[(y_{1i} - d_i \alpha - x_i'\beta)^2] \ : \ \beta_j = 0, \forall j \not \in \hat I \}.
$$
The set $\hat I$ can contain other variables with names $\hat I_3$ that the analyst may think are important
for ensuring robustness.  Thus, $\hat I =  \hat I_1 \cup \hat I_2 \cup \hat I_3$; let $\hat s = |\hat I|$
and $\hat s_j = |\hat I_j|$ for $j =1,2,3$.

\textbf{Condition ASTE}. \textit{(i) The data $(y_{1i}, d_i, z_i), i=1,...,n$, obeys model
(\ref{eq: PL1})-(\ref{Def: errors})  for each $n$, and $x_i = P(z_i)$ is a dictionary
of transformations of $z_i$. (ii) The parameter values $\sigma^2_v$ and $\sigma^2_{\zeta}$ are
bounded from above by $\bar \sigma$ and away from zero, uniformly in $n$, and $|\alpha_0|$
is bounded uniformly in $n$. (iii) Regressor values $x_i, i=1,...,n$, obey the normalization condition $\En[x^2_{ij}]=1$
for all $j \in \{1,...,p\}$ and sparse eigenvalue condition SE. (iv) There exists $s \geq 1$ and $\beta_{m0}$ and $\beta_{g0}$ such that
\begin{eqnarray}
&&m(z_i) = x_i' \beta_{m0} + r_{mi},  \ \ \|\beta_{m0}\|_0 \leq s, \ \  \{ \En[r_{mi}^2]\}^{1/2} \leq K  \bar \sigma \sqrt{s/n}, \\
&&g(z_i) = x_i' \beta_{g0} + r_{gi},  \ \ \|\beta_{g0}\|_0 \leq s, \ \  \{ \En[r_{gi}^2]\}^{1/2} \leq K  \bar \sigma \sqrt{s/n},
\end{eqnarray}
where $K$ is an absolute constant, independent of $n$, but all other parameter values can depend $n$.
(v) $s^2 \log^2 (p\vee n) = o(n)$ and $ \hat s_3 \lesssim  1\vee \hat s_1 \vee \hat s_2$.
 }

\begin{theorem}[Inference on Treatment Effects]\label{theorem:inference} Suppose condition ASTE holds.  The double-Post-Lasso estimator
$\check \alpha$ obeys,
$$
(\sigma_\zeta^2 [\En v_i^2]^{-1})^{-1/2} \sqrt{n} (\check \alpha - \alpha_0) = N(0,1) + o_P(1).
$$
Moreover, the result continues to apply if $\sigma_\zeta^2$ is replaced
by $\hat \sigma_\zeta^2 = \En[(y_{1i} - d_i\check \alpha - x_i'\check \beta)^2](n/(n - \hat s-1))$
and $\En[v_i^2]$ by $\En[\hat v_i^2]= \min_{\beta \in \Bbb{R}^p} \{\En[(d_i - x_i'\beta)^2]: \beta_j =0, \forall j \not \in \hat I \}$.
\end{theorem}

\begin{remark} Theorem \ref{theorem:inference}, derived by the second-named author, shows that the double-Post-Lasso estimator asymptotically achieves the semi-parametric efficiency bound under a set of technical conditions and the following key growth condition:
$ s^2 \log^2 (p\vee n) = o(n).$
This rate condition requires the conditional expectations to be sufficiently smooth  so that a
relatively small number of series terms can be used to approximate them well. As in the case of the IV estimator, this condition can be replaced with the weaker condition that
$ s \log (p\vee n) = o(n)$
by employing a sample splitting method of \citen{FanGuoHao2011}. This is done
in a companion paper, which also deals with a more general setup, covering non-Gaussian,
heteroscedastic disturbances \cite{BCH:PLinference}.$\qed$ \end{remark}

\begin{remark} The post double selection estimator is formulated in response to the inferential non-robustness properties of the post single selection procedures.  The non-robustness of the latter is in line with the uniformity/robustness critique developed by \citen{Potscher2009}. The post double selection procedure developed here is in part motivated as a constructive response to this uniformity critique. The need for such constructive response was stressed by \citen{Hansen2005}.  The goal here is to produce an inferential method which gives useful confidence intervals that are as robust as possible. Indeed, this robustness is captured by the fact that Theorem \ref{theorem:inference} permits  the data-generating process (dgp) to change with $n$, as explicitly stated in the Notation section.  Thus conclusions of the theorem are valid for a wide variety of sequences of dgps. However, while this construction partly addresses the uniformity critique, it does not achieve ``full" uniformity, that is, it does not achieve validity over \emph{all} potential sequences of dgps.  However, we should not interpret this as a deficiency, if the potential sequences causing invalidity are thought of as implausible or unlikely (see \citen{gine:nickl}).  Finally,  it would be desirable to have a \emph{useful} procedure that is valid under \emph{all} sequences of dgps, but such a procedure does not exist.
\qed \end{remark}

\subsection{Monte Carlo Example: Partially Linear Models}

In this section, we compare the estimation strategies proposed above  in the following model:
\begin{equation}\label{ModelMCPLM}y_i = d_i'\alpha_0 + \tilde x_i'\beta_0  + \zeta_i, \ \  \zeta_i \sim N(0,\sigma_\zeta^2)\end{equation}
where the covariates $\tilde x \sim N(0,\Sigma)$, $\Sigma_{kj} = (0.5)^{|j-k|}$, and
\begin{equation}\label{ModelMCPLMd} d_i =  \tilde x_i'\eta_0  + v_i, \ \  v_i \sim N(0,\sigma^2_v)\end{equation}
with $\sigma_\zeta = \sigma_v =1$, and $\sigma_{\zeta v}=0$.
The dimension $p$ of the covariates $x$ is $200$, and the sample size $n$ is $100$. We set $\alpha_0 =1$ and
\begin{eqnarray*}
\beta_0  & = &  \left (1,\frac{1}{2},\frac{1}{3},\frac{1}{4},\frac{1}{5},  \ 0, \ 0, \ 0, \ 0,  \ 0,  \ 1,  \frac{1}{2},\frac{1}{3},\frac{1}{4},\frac{1}{5},0,\ldots,0 \right)', \\
\eta_0 & = & \left(1,\frac{1}{2},\frac{1}{3},\frac{1}{4},\frac{1}{5},\frac{1}{6},\frac{1}{7},\frac{1}{8},\frac{1}{9}, \frac{1}{10}, \ 0,\ldots \ldots \ldots \ldots \ldots ,0 \right)'.
 \end{eqnarray*}
We set  $\lambda$ according to the $X$-dependent rule with $1-\conflvl = .95$.  For each repetition we draw new $x$'s, $\zeta$'s and $v$'s.

We summarize the inference performance of these methods in  Table \ref{Fig:Cover} which illustrates mean bias, standard deviation, and rejection probabilities of 95\% confidence intervals. As
we had expected, Lasso and Post-Lasso exhibit a large mean bias which dominates the estimation error and results in poor performance of conventional inference methods.  On the other hand, the Indirect Post-Lasso has a small bias relative to estimation error but is substantially more variable than double-Post-Lasso
and produces a conservative test, a test with size much smaller than the nominal level.
Notably, the double-Post-Lasso provides coverage that is close to the promised $5\%$ level and has the smallest
mean bias and  standard deviation.  


{\small \begin{table}[h!]\title{\textbf{Partial Linear Model Simulation Results}} \\

\begin{tabular}{lcccc}
Estimator & Mean Bias & Std. Dev. & rp(0.05)\\
\hline
Lasso      & 0.644 & 0.093 & 1.000 \\
Post-Lasso & 0.415 & 0.209 & 0.877  \\
Indirect Post-Lasso & 0.0908 & 0.194 & 0.004  \\
Double selection  & -0.0041  & 0.111 & 0.054 \\
Double selection Oracle & 0.0001 & 0.110 & 0.051 \\
Oracle & -0.0003 & 0.100  & 0.044 \\
\hline
\end{tabular} \caption{\footnotesize Results are based on 1000 simulation replications of the partially linear model (\ref{ModelMCPLM}) where $p=200$ and $n=100$. We report mean bias (Mean Bias), standard deviation (Std. Dev.), and rejection frequency for 5\% level tests (rp(.05)) for the four estimators described in Section 7.1.}\label{Fig:Cover}
\end{table}}

\section{Empirical Examples.}\label{Sec:Empirical}

In this section, we illustrate the performance of sparse methods in two empirical examples.
In the first, we revisit the classic \citen{AK1991}'s instrumental variables estimation of the returns to schooling.  In this example, there are many instruments which can potentially be used in forming the IV estimator and there are concerns about the potential biases and inferential problems introduced from using many instruments.  Our results show that sparse methods can be effectively used to alleviate these concerns.  The second example concerns the use of $\ell_1$-penalized methods to select control variables for growth regressions in which there are many possible country level controls relative to the number of countries.  Using Square-root Lasso to select control variables, we find that there is evidence in favor of the hypothesis of convergence.

\subsection{Angrist and Krueger Example with 1530 instruments}

We consider the \citen{AK1991} model $$
\begin{array}{lll}
 y_{1i}  & = \theta_1 y_{2i} +
w_i'\gamma + \zeta_i, &  \Ep[\zeta_i|w_i, z_i] = 0,\\
 y_{2i}  & = z_i'\beta + w_i'\delta +
v_i,  &  \Ep[v_i|w_i, z_i] = 0,
\end{array}
 $$
where  $y_{1i}$ is the log(wage) of individual $i$,  $y_{2i}$ denotes education, $w_i$ denotes a vector of control variables, and $z_i$ denotes a vector of instrumental variables that affect education but do not directly affect the wage. The data were drawn from the 1980 U.S. Census and consist of 329,509 men born between 1930 and 1939.  In this example, $w_i$ is a set of 510 variables: a constant, 9 year-of-birth dummies, 50 state-of-birth dummies, and 450 state-of-birth $\times$ year-of-birth interactions.  As instruments, we use three quarter-of-birth dummies and interactions of these quarter-of-birth dummies with the set of state-of-birth and year-of-birth controls in $w_i$ giving a total of 1530 potential instruments.  \citen{AK1991} discusses the endogeneity of schooling in the wage equation and provides an argument for the validity of $z_i$ as instruments based on compulsory schooling laws and the shape of the life-cycle earnings profile.  We refer the interested reader to \citen{AK1991} for further details.  The coefficient of interest is $\theta_1$, which summarizes the causal impact of education on earnings.

There are two basic options for estimating $\theta_1$ that have been used in the literature: one uses just the three basic quarter-of-birth dummies and the other uses 180 instruments corresponding to the three quarter-of-birth dummies and their interactions with the 9 main effects for year-of-birth and 50 main effects for state-of-birth.  It is commonly-held that using the set of 180 instruments results in 2SLS estimates of $\theta_1$ that have a substantial bias, while using just the three quarter-of-birth dummies results in an estimator with smaller bias but a large variance; see, e.g., \citen{hhn:weakiv}.  Another approach uses the 180 instruments and the Fuller estimator \cite{fuller}  (FULL) with an adjustment for the use of many instruments.  Of course, using sparse methods for the first-stage estimation offers another option that could be used in place of any of the aforementioned approaches.

{\small \begin{table}[!h]\title{Estimates of the Returns to Schooling in the Angrist-Krueger Data}
\begin{tabular}{lcccc}
\hline
\hline
\multicolumn{1}{c}{Number of} &	\multicolumn{4}{c}{ } \\
\multicolumn{1}{c}{Instruments} &	2SLS Estimate &	2SLS Std. Error	& 	 Fuller Estimate  &	 Fuller Std. Error \\
\hline
3 & 0.1079 & 0.0196 & 0.1087 & 0.0200 \\
180 & 0.0928 & 0.0097 & 0.1063 & 0.0143 \\
1530 & 0.0712 & 0.0049 & 0.1019 & 0.0422 \\
\hline
\multicolumn{5}{c}{Lasso - Iterated}\\
\hline
1 & 0.0862 & 0.0254 & & \\
\hline
\multicolumn{5}{c}{Lasso - 10-Fold Cross-Validation} \\
\hline
12 & 0.0982 & 0.0137 & 0.0997 & 0.0139 \\
\hline
\hline
\hline
\multicolumn{5}{c}{Sup-Score/Inverse Lasso $95\%$ Confidence Interval} \\
\multicolumn{1}{c}{Number of} &	\multicolumn{4}{c}{ } \\
\multicolumn{1}{c}{Instruments} &	Center of CI &   Quasi Std. Error  &   Confidence Interval    &  \\
\hline
3 & .100 & 0.0255  & (0.05,0.15) \\
180 & .110 &  0.0459  & (0.02,0.20) \\
1530 & .095 & 0.0689 & (-0.04,0.23) \\
\hline
\hline
\end{tabular}\caption{{\footnotesize This table reports estimates of the returns-to-schooling parameter in the Angrist and Krueger 1991 data for different sets of instruments.  The columns 2SLS and 2SLS Std. Error give the 2SLS point estimate and associated estimated standard error, and the columns Fuller Estimate and Fuller Std. Error give the Fuller point estimate and associated estimated standard error.  We report Post-Lasso results based on instruments selected using the plug-in penalty described in Section 3.1 (Lasso - Iterated) and based on instruments selected using a penalty level chosen by 10-Fold Cross-Validation (Lasso - 10-Fold Cross-Validation).  For the Lasso-based results, Number of Instruments is the number of instruments selected by Lasso.}}
\end{table}}

Table 5 presents estimates of the returns to schooling coefficient using 2SLS and FULL\footnote{We set the user-defined choice parameter in the Fuller estimator equal to one which results in a higher-order unbiased estimator.} and different sets of instruments.  Given knowledge of the construction of the instruments, the first three rows of the table correspond to the natural groupings of the instruments into the three main quarter of birth effects, the three quarter-of-birth dummies and their interactions with the 9 main effects for year-of-birth and 50 main effects for state-of-birth, and the full set of 1530 potential instruments.  The remaining two rows give results based on using Lasso to select instruments with penalty level given by the simple plug-in rule in Section 3 or by 10-fold cross-validation.  Using the plug-in rule, Lasso selects only the dummy for being born in the fourth quarter; and with the cross-validated penalty level, Lasso selects 12 instruments which include the dummy for being born in the third quarter, the dummy for being born in the fourth quarter, and 10 interaction terms.  The reported estimates are obtained using Post-Lasso.

The results in Table 5 are interesting and quite favorable to the idea of using Lasso to do variable selection for instrumental variables.  It is first worth noting that with 180 or 1530 instruments, there are modest differences between the 2SLS and FULL point estimates that theory as well as evidence in \citen{hhn:weakiv} suggests is likely due to bias induced by overfitting the 2SLS first-stage which may be large relative to precision.  In the remaining cases, the 2SLS and FULL estimates are all very close to each other suggesting that this bias is likely not much of a concern.  This similarity between the two estimates is reassuring for the Lasso-based estimates as it suggests that Lasso is working as it should in avoiding overfitting of the first-stage and thus keeping bias of the second-stage estimator relatively small.

For comparing standard errors, it is useful to remember that one can regard Lasso as a way to select variables in a situation in which there is no \textit{a priori} information about which of the set of variables is important; i.e. Lasso does not use the knowledge that the three quarter of birth dummies are the ``main'' instruments and so is selecting among 1530 \textit{a priori} ``equal'' instruments.  Given this, it is again reassuring that Lasso with the more conservative plug-in penalty selects the dummy for birth in the fourth quarter which is the variable that most cleanly satisfies \citen{AK1991}'s argument for the validity of the instrument set.  With this instrument, we estimate the returns-to-schooling to be .0862 with an estimated standard error of .0254.  The best comparison is FULL with 1530 instruments which also does not use any \textit{a priori} information about the relevance of the instruments and estimates the returns-to-schooling as .1019 with a much larger standard error of .0422.  One can be less conservative than the plug-in penalty by using cross-validation to choose the penalty level.  In this case, 12 instruments are chosen producing a Fuller point estimate (standard error) of .0997 (.0139) or 2SLS point estimate (standard error) of .0982 (.0137).  These standard errors are smaller than even the standard errors obtained using information about the likely ordering of the instruments given by using 3 or 180 instruments where FULL has standard errors of .0200 and .0143 respectively.  That is, Lasso finds just 12 instruments that contain nearly all information in the first stage and, by keeping the number of instruments small, produces a 2SLS estimate that likely has relatively small bias.
We believe that these empirical results are reliable.   In particular,  we note that the first stage $F$ statistic
on the selected 12 instruments is approximately $20$;  our computational experiments in the previous section employ designs with $F=10$ and $F=40$ to show that
this method works well for both estimation and inference purposes.


As a final check, we report the 95\% confidence interval obtained from the Sup-Score test of Section 5.2 based on the three natural groupings of 3, 180, and 1530 instruments.  This test is robust to weak or non-identification and is simple to implement.  For the three different sets of instruments, we obtain intervals that are much wider but roughly in line with the intervals discussed above.  We note that our preferred method from the simulation section only makes use of the Sup-Score test when no instruments are selected, does a good job at controlling size in the simulation, and is more powerful than the Sup-Score test when the instruments contain signal about the endogenous variable.  Using this procedure would lead us to use the much more precise IV-Lasso results.

Overall, these results demonstrate that Lasso instrument selection is feasible and produces sensible and what appear to be relatively high-quality estimates in this application.  The results from the Lasso-based IV estimators are similar to those obtained from other leading approaches to estimation and inference with many-instruments and do not require \textit{ex ante} information about which are the most relevant instruments.  Thus, the Lasso-based IV procedures should provide a valuable complement to existing approaches to estimation and inference in the presence of many instruments.


\subsection{Growth Example}

In this section, we consider variable selection in an international economic growth example. We use the \citen{BarroLee1994} data consisting of a panel of 138 countries for the period of 1960 to 1985. We consider the national growth rates in GDP per capita as the dependent variable.  In our analysis, we consider a model with $p=62$ covariates which allows for a total of $n=90$ complete observations.  Our goal here is to provide estimates which shed light on the convergence hypothesis discussed below by selecting controls from among these covariates.\footnote{We can compare our results to those obtained in other standard models in the growth literature such as \cite{BarroSala1995,KoenkerMachado1999}.}

One of the central issues in the empirical growth literature is the estimation of the effect of an initial (lagged) level of GDP per capita on the growth rates of GDP per capita. In particular, a key prediction from the classical Solow-Swan-Ramsey growth model is the hypothesis of convergence which states that poorer countries should typically grow faster than richer countries and therefore should tend to catch up with the richer countries over time. This hypothesis implies that the effect of a country's initial level of GDP on its growth rate should be negative. As pointed out in Barro and Sala-i-Martin \citeyear{BarroSala1995}, this hypothesis is rejected using a simple bivariate regression of growth rates on the initial level of GDP. (In our case, regression yields a statistically insignificant coefficient of $.00132$.) In order to reconcile the data and the theory, the literature has focused on estimating the effect \textit{conditional} on characteristics of countries.  Covariates that describe such characteristics can include variables measuring education and science policies, strength of market institutions, trade openness, savings rates and others; see \cite{BarroSala1995}.  The theory then predicts that the effect of the initial level of GDP on the growth rate should be negative among otherwise similar countries.  

Given that the number of covariates we can condition on is comparable to the sample size, covariate selection becomes an important issue in this analysis; see \citen{OneMillion}, \citen{TwoMillion}, \citen{Sala-i-MartinDoppelhoferMiller2004}.  In particular, previous findings came under severe criticisms for relying upon \textit{ad hoc} procedures for covariate selection; see, e.g., \citen{OneMillion}.   Since the number of covariates is high, there is no simple way to resolve the model selection problem using only standard tools. Indeed the number of possible lower-dimensional model is very large, though see \citen{OneMillion}, \citen{TwoMillion} and \citen{Sala-i-MartinDoppelhoferMiller2004} for attempts to search over millions of these models. Here we use $\ell_1$-penalized methods to attempt to resolve this important issue.

We first present results for covariate selection using the different methods discussed in Section \ref{Sec:Treatment}:
(a) a simple Post-Square-root-Lasso method which uses controls selected from applying the Square-root-Lasso
to select controls in the regression of growth rates on log-GDP and other controls, and (b) the Post-double-selection
method, which uses the controls selected by Square-root-Lasso in the regression of log-GDP on other controls and in the regression of growth rates on other controls.  These were all based on Square-root Lasso to avoid the estimation of $\sigma$.  We present the model selection results in Table \ref{Table:Growth}.

{\small
\begin{table}\begin{center}{ \bf Model Selection Results for the International Growth Regressions \\ Real GDP per capita (log) is included in all models }
\renewcommand{\arraystretch}{1}
 \begin{tabular}{ccc}
\hline Selection Method    
 &  & Additional Variables Selected \\ \hline 
Square-root Lasso  &  &  Black Market Premium (log)  \\
\rowcolor[gray]{0.9}  Double selection &  &  Terms of trade shock \\ 
\rowcolor[gray]{0.9}                  &  & Infant Mortality Rate (0-1 age)\\
\rowcolor[gray]{0.9}           &  & Female gross enrollment for secondary education \\
\rowcolor[gray]{0.9}                  &  & Percentage of ``no schooling" in the female population \\ 
\rowcolor[gray]{0.9}                  &  & Percentage of ``higher school attained" in the male population \\ 
\rowcolor[gray]{0.9}                  &  & Average schooling years in the female population over the age of 25 \\ 
\\
\hline
\\
\end{tabular}\caption{\footnotesize The controls selected by different methods.}\label{Table:Growth}
\end{center}
\end{table}}

Square-root Lasso applied to the regression of growth rates on log-GDP and other controls selected only one control, the log of the black market premium which characterizes trade openness.  The double selection method
selected infant mortality rate, terms of trade shock, and several education variables (female gross enrollment for secondary education, percentage of ``no schooling" in the female population, percentage of ``higher school attained" in male population, and average schooling years in female population over the age of 25) to forecast log-GDP but no additional controls were selected to forecast growth.
We refer the reader to \citen{BarroLee1994} and \citen{BarroSala1995} for a complete definition and discussion of each of these variables.

We then proceeded to construct confidence intervals for the coefficient on initial GDP based on each set of selected variables.   We also report estimates of the effect of initial GDP in a model which uses the set of controls obtained from the double-selection procedure and additionally includes the log of the black market premium. We expressly allow for such amelioration strategy in our formal construction
of the estimator. Table \ref{Table:CondidenceInterval} shows these results.
We find that in all these models the linear regression coefficients on the initial level of GDP are negative.  In addition, zero is excluded from the 90\% confidence interval in each case.  These findings support the hypothesis of (conditional) convergence derived from
the classical Solow-Swan-Ramsey growth model.  The findings also agree with and thus support the previous findings reported in   \citen{BarroSala1995} which relied on ad-hoc reasoning for covariate selection.


{\small \begin{table}
\begin{center}
{\bf Confidence Intervals after Model Selection \\ for the International Growth Regressions}
\begin{tabular}{lcccc}
\\
\hline
 & & \multicolumn{2}{c}{Real GDP per capita (log)} \\  Method & & Coefficient & $90\%$ Confidence Interval \\
\hline
Post Square-root Lasso & & $-0.0112$      & $[-0.0219,  -0.0007]$\\
Post Double selection && $-0.0221$ & $[-0.0437,-0.0005]$\\
Post Double selection (+ Black Market Premium) && $-0.0302$  & $ [ -0.0509, -0.0096 ]$\\

\hline

\end{tabular}\caption{\footnotesize The table above displays the coefficient and a $90\%$ confidence interval associated with each method. The selected models are displayed in Table \ref{Table:Growth}.}\label{Table:CondidenceInterval}\end{center}
\end{table}}

\section{Conclusion}
There are many situations in economics where a researcher has access to data with a large number of covariates.  In this article, we have presented results for performing analysis of such data by selecting relevant regressors and estimating their coefficients using $\ell_1$-penalization methods.  We gave special attention to the instrumental variables model and the partially linear model, both of which are widely used to estimate structural economic effects.  Through simulation and empirical examples, we have demonstrated that $\ell_1$ penalization methods may be usefully employed in these models and can complement tools commonly employed by applied researchers.

Of course, there are many avenues for additional research.  The use of $\ell_1$-penalization is only one method of performing estimation with high-dimensional data.  It will be interesting to consider and understand the behavior of other methods (e.g. \citen{HHS2008}, \citen{FanLi2001},
\citen{zhang:concave}, \citen{fan:liao}) for estimating structural economic objects.  In addition, extending HDS models and methods to other types of economic models beyond those considered in this article will be interesting.  An important problem in economics is the analysis of high-dimensional data in which there are many weak signals within the set of variables considered in which case the sparsity assumption may provide a poor approximation. The sup-score test presented in this article offers one approach to dealing with this problem, but further additional research dealing with this issue seems warranted. It would also be interesting to consider efficient use of high-dimensional data in cases in which scores are not independent across observations which is a much-considered case in economics.  Overall, we believe the results in this article provide useful tools for applied economists but that there are still substantial and interesting topics in the use of high-dimensional economic data that warrant further investigation.

\appendix

\small
\section{Iterated Estimation of the Noise Level $\sigma$}\label{Sec:EstSigma}

In the case of Lasso, the penalty levels (\ref{Def:LambdaLASSOboound}) and (\ref{Def:LambdaLASSO}) require the practitioner to fill in a value for $\sigma$.
Theoretically, any upper bound on $\sigma$ can be used and the standard approach in the literature is to
use the conservative estimate $\bar \sigma = \sqrt{\text{Var}_n[y_i]}:= \sqrt{\En\[(y_i - \bar y)^2\]}$, where  $\bar y = \En[y_{i}]$. Unfortunately, in various examples we found that this approach leads to overpenalization. Here we briefly discuss iterative procedures to estimate $\sigma$ similar to the ones described in \citen{BC-LectureNotes}. Let  $I_0$ be a set of regressors that is included in the model.  Note that $I_0$ is always non-empty since it will always include the intercept. Let $\bar\beta(I_0)$ be the least squares estimator of the coefficients on the covariates associated with $I_0$, and define
$ \hat \sigma_{I_{0}} := \sqrt{\En[ (y_i-x_i'\bar\beta(I_0))^2]}.$

An algorithm for estimating $\sigma$ using Lasso is as follows:
\begin{algorithm}[Estimation of $\sigma$ using Lasso iterations] For a positive number $\psi$, set $\hat \sigma_0 = \psi \hat\sigma_{I_{0}}$.  Set $k = 0$, and specify a small constant $\nu \geq 0$ as a tolerance level and
a constant $K>1$ as an upper bound on the number of iterations.   (1)  Compute the Lasso estimator $\widehat \beta$ based on $\lambda = 2c \hat \sigma_k \Lambda(1-\gamma|X)$.(2) Set $\hat \sigma_{k+1}^2 = \widehat Q(\widehat \beta).$ (3) If $| \hat \sigma_{k+1} - \hat \sigma_{k}| \leqslant \nu$ or $k> K$, report $\hat\sigma = \hat \sigma_{k+1}$;
otherwise set $k \leftarrow k+1 $ and go to (1).
\end{algorithm}

Similarly, an algorithm for estimating $\sigma$ using Post-Lasso is as follows:
\begin{algorithm}[Estimation of $\sigma$ using Post-Lasso iterations] For a positive number $\psi$, set $\hat \sigma^0 = \psi\hat\sigma_{I_{0}}$.  Set $k = 0$, and specify a small constant $\nu \geq 0$ as a tolerance level and
a constant $K>1$ as an upper bound on the number of iterations. (1)  Compute the Post-Lasso estimator $\widetilde \beta$ based on $\lambda = 2c \hat \sigma_k \Lambda(1-\gamma|X)$. (2) For $\widehat s = \|\widetilde \beta\|_0 = |\widehat T|$ set $\hat \sigma_{k+1}^2 =   \widehat Q(\widetilde \beta) \cdot n/(n-\widehat s).$ (3) If $| \hat \sigma_{k+1} - \hat \sigma_{k}| \leqslant \nu$ or $k > K$, report $\hat\sigma = \hat \sigma_{k+1}$; otherwise, set $k \leftarrow k+1 $ and go to (1).
\end{algorithm}

\begin{remark}
We note that we employ the standard degree-of-freedom correction with $\widehat s = \|\widetilde \beta\|_0 = |\widehat T|$ when using Post-Lasso (Algorithm 2).  No additional correction is necessary when using Lasso (Algorithm 1) since the Lasso estimate is already sufficiently regularized. We note that the sequence $\widehat \sigma_k$, $k\geqslant 2$, produced by Algorithm 1 is monotone and that the estimates $\widehat \sigma_k$, $k\geqslant 1$, produced by Algorithm 2 can only assume a finite number of different values. \citen{BC-LectureNotes} and \citen{BC-PostLASSO} provide theoretical analysis for $\psi = 1$. In preliminary simulations with coefficients that were not well separated from zero, we found that $\psi = 0.1$ worked better than $\psi = 1$ by avoiding unnecessary overpenalization in the first iteration. \qed
\end{remark}

\section{Proof of Theorem \ref{Thm:IV}}

Step 1. Recall that $A_i = (f(z_i),w_i')'$ and $d_i = (y_{2i},w_i')'$ for $i=1,\ldots,n$.   Let $X=[x_1,\ldots,x_n]'$,
$A = [A_1,...,A_n]'$, $D= [d_1,...,d_n]'$, $W=[w_1,...,w_n]'$, $f=[f(z_1),...,f(z_n)]'$, $Y_2 = [y_{21},...,y_{2n}]'$,
$V =[v_1,...,v_n]'$, and $\zeta=[\zeta_1,...,\zeta_n]'$.  We have that
\begin{eqnarray*}
\sqrt{n}(\widehat \alpha^* - \alpha) &= &  [\widehat A' D/n]^{-1}  \widehat A'\zeta/\sqrt{n} =  \[Q_n + o_P(1)\]^{-1} \left( A'\zeta/\sqrt{n} + o_P(1) \right)
 \end{eqnarray*}
where by Steps 3 and 4 below:
\begin{eqnarray}
& \widehat A'D/n = A'D/n + o_{P}(1) =  Q_n + o_P(1) \label{eq: to show 1} \\
&  \widehat A'\zeta/\sqrt{n} = A'\zeta/\sqrt{n} + o_P(1)\label{eq: to show 2}.
 \end{eqnarray}
Moreover, by the assumption on $\sigma_\zeta$ and $Q_n$, $\textrm{Var}(A'\zeta/\sqrt{n} ) = \sigma^2_\zeta Q_n$ has eigenvalues bounded away from zero and bounded from above, uniformly in $n$. Therefore,
$
\sqrt{n}(\widehat \alpha^* - \alpha_0) = Q^{-1}_n  A'\zeta/\sqrt{n} + o_P(1),
$
and  $Q^{-1}_n   A'\zeta/\sqrt{n}$ is a vector distributed as normal with mean zero and covariance
$\sigma^2_\zeta Q^{-1}_n$.  This verifies the main claim of the theorem.

Step 2.  This is an auxiliary step where we note that conditions of the theorem imply
by Markov inequality:
\begin{eqnarray*}
 & & f'f/n+  {\rm tr } (W'W/n) = {\rm tr } (A'A/n)  = {\rm tr }(Q_n)\lesssim 1,  \\
  & &  \|D'\zeta/n\| \leq |V'\zeta/n| + \|A'\zeta/n\| \lesssim_P  \sigma_{\zeta v}+ 1/\sqrt{n},\\
 & & \|A'V/n\|^2  = |f'V/n|^2 + \|W'V/n\|^2 \lesssim_P 1/n,   \\
 & & \|D/\sqrt{n}\| \leq \|V/\sqrt{n}\| + \|A/\sqrt{n}\| \lesssim_P 1.
\end{eqnarray*}

Step 3.  To show (\ref{eq: to show 1}), note that $\widehat A - A = (\widehat f' - f', 0')'$. Thus,
 \begin{eqnarray*}
\| \widehat A'D/n - A'D/n \| = | (\widehat f - f)'Y_2/n| &\leq&   \sqrt{ (\widehat f - f)' (\widehat f - f)/n}\sqrt{Y_2'Y_2/n} = o_P(1)
 \end{eqnarray*}
since $\sqrt{Y_2'Y_2/n} \lesssim_P 1$  by Markov inequality, and
$\sqrt{ (\widehat f - f)' (\widehat f - f)/n} = o_{P}(1)$ by Theorems \ref{corollary1:rate} or \ref{corollary3:postrate}.
Next, since $f'V/n = o_P(1)$ and $W'V/n=o_P(1)$ by Step 2,
note that $A'D/n = A'A/n + o_P(1) = Q_n + o_P(1)$.

Step 4.  To show (\ref{eq: to show 2}), note that
 \begin{eqnarray*}
  \|(\widehat A- A)' \zeta/\sqrt{n}\| &  =  | (\widehat f - f)' \zeta/\sqrt{n}|  =  | (X(\widehat \beta - \beta_{0}))'\zeta/\sqrt{n} +  (f- X\beta_0)'\zeta/\sqrt{n}  |  \\
&   \leq
\left\| X'\zeta/\sqrt{n}  \right\|_\infty
  \|\widehat \beta - \beta_{0}\|_{1}   +   |(f- X\beta_0)'\zeta/\sqrt{n}  | \to_P 0.
 \end{eqnarray*}
 This follows because the first term is of order $\sqrt{\log (p\vee n)} \sqrt{ s^2 \log (p\vee n) /n } \to 0$  by
 conditions of the theorem; the order follows because
$ \left\| X'\zeta/\sqrt{n}  \right\|_\infty \lesssim_P \sqrt{ \log (p\vee n)}$ by (\ref{Eq:GaussIneq}), and
$ \|\widehat \beta - \beta_{0}\|_{1} \lesssim_P \sqrt{ [s^2 \log (p\vee n)] /n }$ by Theorems \ref{corollary1:rate} and \ref{corollary3:postrate} since
$ \|\widehat \beta - \beta_{0}\|_{1} \leq \sqrt{s+\hat s}\|\widehat \beta - \beta_{0}\|\lesssim_P \sqrt{ [s^2 \log (p\vee n)] /n }$ under condition SE and $\hat s \lesssim_P s$.
On the other hand, the second term converges to zero in probability by Markov
inequality, because the expectation of $|(f- X\beta_0)'\zeta/\sqrt{n}|^2$ is of order $\sigma^2_{\zeta}c^2_s \to 0$.

Step 5. This step establishes
consistency of the variance estimator.
Since $\sigma^2_\zeta$ and the eigenvalues of $Q_n$ are bounded away from zero
and from above uniformly in $n$, it suffices to show $\hat \sigma^2_\zeta - \sigma^2_\zeta \to_P 0$
and  $\widehat A'\widehat A/n - Q_n \to_P 0$. Indeed,
$\hat\sigma^2_\zeta = \|\zeta-D(\widehat\alpha^*-\alpha_0)\|^2/n =
\|\zeta\|^2/n +2 \zeta'D(\alpha_0-\widehat\alpha^*)/n
+ \|D(\alpha_0-\widehat\alpha^*)\|^2/n$ so that
$\|\zeta\|^2/n - \sigma^2_\zeta \to_P 0$ by Chebyshev inequality since $\max_i \Ep[\zeta_i^4]$ is bounded
uniformly in $n$, and the remaining
terms converge to zero in probability since $\widehat \alpha^* - \alpha_0 \to_P
0$, $\|D'\zeta/n\| \lesssim_P 1$ by Step 2. Next, note that
$$ \| \widehat A'\widehat A/n - A'A/n \| = \|
 A'(\widehat A-A)/n + (\widehat A-A)'A/n +
(\widehat A-A)' (\hat A-A)/n \|
$$
which is bounded up to a constant by
$
(\|\widehat A-A\|/\sqrt{n})(\|A\|/\sqrt{n}) + \|\widehat A-A\|^2/n \to_P 0
$
since $\|\widehat A-A\|^2/n = \|\widehat f-f\|^2/n=o_P(1)$ by  Theorems \ref{corollary1:rate} or \ref{corollary3:postrate}, and  $\|A\|^2/n \lesssim_P 1$ holding by Step 2.\qed

\section{Proof of Theorem \ref{Thm:WIV}}
 Step 1. When $a = \alpha_1$ we have that
$$
\Lambda_{\alpha_1}  = \max_{1 \leq j \leq p} \frac{n|\En[\tilde \epsilon_i \tilde x_{ij}] |}{\sqrt{\En[\tilde \epsilon_i^2 \tilde x_{ij}^2]}} =
\max_{1 \leq j \leq p} \frac{n|\En[\tilde g_i \tilde x_{ij}] |}{\sqrt{\En[\tilde g_i^2 \tilde x_{ij}^2]}}
$$
so claim (1) follows from the definition of quantile and from the continuity of the distribution  of $\Lambda_{\alpha_1}$.

Step 2. To establish claim (2), we note that
$$
n \En[\tilde g_i \tilde x_{ij}] = n \En[g_i \tilde x_{ij}] = \sqrt{n} \mathcal{N}_j \sqrt{\En[\tilde x_{ij}^2]} = \sqrt{n} \mathcal{N}_j,
$$
where $\mathcal{N}_j \sim N(0,1)$ for each $j$.  Since for $\hat \mu_g = (\En[w_i w_i'])^{-1} \En[w_i g_i]$ we have
$\|\hat \mu_g \| \lesssim_P  1/\sqrt{n}$ by
the assumed boundedness of $\|(\En[w_i w_i'])^{-1}\|$ and boundedness of $\|w_i\|$,
we conclude that $\max_{i \leq n } |w_i' \hat \mu_g| \lesssim_P 1/\sqrt{n}$, so that
$$
|\sqrt{\En [\tilde g_i^2 \tilde x^2_{ij}]} -\sqrt{\En [g_i^2 \tilde x^2_{ij}]}| \leq \sqrt{ \En{ [(w_i'\hat \mu_g)^2 \tilde x^2_{ij} ]} } \lesssim_P
n^{-1/2}\sqrt{\En [\tilde x^2_{ij}]},
$$
uniformly in $j \in \{1,...,p\}$, using the triangular inequality and the decomposition $\tilde g_i = g_i - w_i' \hat \mu_g$.
Moreover, using the Bernstein-type inequality in Lemma 5.15 of \citen{vandeGeer:book}, we can conclude that
$$
|\En[g_i^2 \tilde x_{ij}^2] - \En[\tilde x_{ij}^2] | \lesssim_P \sqrt{ (\log p)/ n},
 $$
uniformly in $j \in \{1,...,p\}$.  Hence since $\En[\tilde x_{ij}^2]=1$ by the normalization assumption, we conclude
that with probability approaching 1,
$$
\Lambda_{\alpha_1} \leq \max_{1 \leq j \leq p} c n|\En[ g_i \tilde x^2_{ij}] |/\sqrt{\En[\tilde x^2_{ij}]} =  \max_{1 \leq j \leq p} c \sqrt{n} |\mathcal{N}_j|
$$
and the  claim (2) follows by the union bound and standard tail properties of $N(0,1)$.

Step 3. To show claim (3) we  note that using triangular and other elementary inequalities:
\begin{eqnarray*}
\Lambda_a & = &  \max_{1 \leq j \leq p} \left | \frac{ n | \En[ (\tilde \epsilon_i - (a-\alpha_1) \tilde y_{2i}) \tilde x_{ij}] }{ \sqrt{
\En[(\tilde \epsilon_i - (a-\alpha_1) \tilde y_{2i})^2 \tilde x^2_{ij}] }} \right| \\
& \geq &
\max_{1 \leq j \leq p} \left | \frac{ |a-\alpha_1| n | \En[ \tilde y_{2i}\tilde x_{ij}]| }{ \sqrt{
\En[\tilde \epsilon^2_i\tilde x^2_{ij}]} +  |a-\alpha_1| \sqrt{
\En[ \tilde y_{2i}^2 \tilde x^2_{ij}] }}   \right| - \Lambda_{\alpha_1}
\end{eqnarray*}
The first term is bounded below by, with probability approaching 1,
$$
c^{-1} \max_{1 \leq j \leq p} \frac{|a- \alpha_1| | n \En[ \tilde y_{2i} \tilde x_{ij} ]|}{\sigma_{\zeta} \sqrt{\En[\tilde x_{ij}^2]} + |a- \alpha_1|\sqrt{\En[\tilde y_{2i}^2 \tilde x_{ij}^2]}},
$$
by Step 2 for some $c>1$,  and $\Lambda_{\alpha_1} \lesssim_P \sqrt{ n \log p}$ by Step 2.  Hence
for any constant $C$, with probability converging to 1, $\Lambda_a - C\sqrt{n \log p} \to + \infty,$
so that Claim (3) immediately follows, since
by Step 2 $\Lambda(1-\gamma|X,W) \lesssim \Lambda(1-\gamma) \lesssim \sqrt{n \log p}$, since
$\gamma \in (0,1)$ is fixed by assumption. \qed

\section{Proof of Theorem \ref{theorem:inference}}  Let me prepare some notation.
I will use the standard matrix notation, namely
$ Y_1 =[y_{11},...,y_{1n}]'$, $X= [x_{1},...,x_{n}]'$, $D=[d_1,...,d_n]'$, $V= [v_1,...,v_n]'$,
$\zeta =[\zeta_1,...,\zeta_n]'$, $m = [m_1,...,m_n]'$ for $m_i = m(z_i)$, $R_m = [r_{m1},...,r_{mn}]'$,
$g = [g_1,...,g_n]'$ for $g_i = g(z_i)$, $R_g = [r_{g1},...,r_{gn}]'$,
and so on. Let $\semin{\hat s}= \semin{\hat s}[\En [x_ix_i']]$. For $A \subset \{1,...,p\}$,
let $X[A] = \{X_j, j \in A\}$,
where $\{X_j, j=1,...,p\}$ are the columns of $X$.  Let
$$\mathcal{P}_A  = X[A](X[A]'X[A])^{-}X[A]'$$ be the projection operator
sending vectors in $\Bbb{R}^n$ onto ${\rm span}[X[A]]$, and let $\mathcal{M}_A = {\rm I}_n - \mathcal{P}_A$ be the
projection onto the subspace that is orthogonal to  ${\rm span}[X[A]]$.  For a vector $Z \in \Bbb{R}^n$, let
$$
\tilde \beta_Z(A) := \arg\min_{b \in \Bbb{R}^p} \|Z- X'b\|^2: \ b_j = 0, \ \forall j \not \in A,
$$
be the coefficient of linear projection of $Z$ onto ${\rm span}[X[A]]$.   If $A = \varnothing$,
interpret $\mathcal{P}_A = 0_n$, and $\tilde \beta_Z = 0_p$.

Step 1.(Main) Write
$
\check \alpha = \[D'\MX D/n\]^{-1}[D'\MX Y_1/n]
$
so that
$$
\sqrt{n}(\check \alpha - \alpha_0) = \[D'\MX D/n\]^{-1}[D'\MX (g + \zeta)/\sqrt{n}] =: ii^{-1} \cdot i.
$$
By Steps 2 and 3, $ii = V'V/n + o_P(1)$ and $i = V'\zeta/\sqrt{n} + o_P(1)$. Since
$V'V/n = \sigma_v^2 + o_P(1)$ by Chebyshev inequality, and $ \sigma^2_{\zeta}$ and $\sigma_v^2$ are bounded from above
and away from zero by assumption, and
$$
V'\zeta/\sqrt{n}  =  [\sigma_\zeta \sqrt{V'V/n}] N(0,1)
$$
conclude that
$$
\sigma^{-1}_{\zeta} (V'V/n)^{1/2} \sqrt{n}(\check \alpha - \alpha_0) = N(0,1) + o_P(1).
$$

Step 2. (Behavior of $i$.) Decompose
\begin{eqnarray}
i = V'\zeta/\sqrt{n} + \underset{=:i_a}{m' \MX g/\sqrt{n}} + \underset{=:i_b}{m'\MX \zeta/\sqrt{n}} + \underset{=:i_c}{V'\MX g/\sqrt{n}} - \underset{=:i_d}{ V'\PX \zeta/\sqrt{n}}.
\end{eqnarray}
First, note that by Steps 4 and 5 and by the growth condition $ s^2\log^2(p\vee n) = o(n)$
$$
|i_a| \leq \sqrt{n} \| m' \MX/\sqrt{n} \| \| g '\MX/ \sqrt{n} \| \lesssim_P \sqrt{n} \sqrt{ [s \log (p\vee n)]^2/n^2 } = o_P(1).
$$
Second, using decomposition $m = X\beta_{m0} + R_m$, bound
$$
|i_b| \leq |R_m'\zeta/\sqrt{n}| + |(\tilde \beta_m(\hat I) - \beta_{m0})'X'\zeta/\sqrt{n}| \lesssim_P \sqrt{ [s \log (p\vee n)]^2/n } = o_P(1),
$$
where $
|R_m'\zeta/\sqrt{n}| \lesssim_P  \sqrt{R_m'R_m/n} \lesssim \sqrt{s/n}
$
by Chebyshev inequality and by assumption ASTE, and
$$
|(\tilde \beta_m(\hat I) - \beta_{m0})'X'\zeta/\sqrt{n}| \leq \|\tilde \beta_m (\hat I) - \beta_{m0}\|_1 \|X'\zeta/\sqrt{n}\|_{\infty} \lesssim_P \sqrt{ [s^2 \log(p\vee n)]/n} \sqrt{\log (p\vee n)},
$$
 $\|\tilde \beta_m (\hat I) - \beta_{m0}\|_1
\leq   \sqrt{\hat s} \|\tilde \beta_m (\hat I) - \beta_{m0}\| \lesssim_P
\sqrt{ [s^2 \log (p\vee n)]/n}$ by Step 4, using that $\hat s \lesssim_P s$ by Theorem 2,  $\|X'\zeta/\sqrt{n}\|_{\infty}\lesssim_P \sqrt{ \log (p\vee n) }$ by the Gaussian
maximal inequality (\ref{Eq:GaussIneq}) and normalization condition on $X$.  Third, using similar reasoning, decomposition $g = X\beta_{g0} + R_g$, and Step 5, conclude
$$
|i_c| \leq |R_g'\zeta| + |(\tilde \beta_g(\hat I) - \beta_{g0})'X'V/\sqrt{n}| \lesssim_P   \sqrt{ [s \log (p\vee n)]^2/n } = o_P(1).
$$
Fourth, using that $\hat s \lesssim_P s$ by Theorem \ref{corollary3:postrate} so that $ 1/\semin{\hat s} \lesssim_P 1$ by condition SE, conclude,
$$
|i_d| \leq |\tilde \beta_V(\hat I)' X'\zeta/\sqrt{n}| \leq \| \tilde \beta_V(\hat I)\|_1 \|X'\zeta/\sqrt{n}\|_{\infty} \lesssim_P  \sqrt{n} \sqrt{ [s \log (p\vee n)]^2/n^2 } = o_P(1),
$$
since $ \| \tilde \beta_V(\hat I)\|_1 \leq \sqrt{\hat s} \|\tilde \beta_V(\hat I)\| \leq$
$\sqrt{\hat s} \| (X[\hat I]'X[\hat I])^{-1} X[\hat I]'V/n\|$ $\leq$
$\sqrt{\hat s} \phi^{-1}_{\min}(\hat s)$  $\sqrt{\hat s}  \| X'V/\sqrt{n}\|_{\infty}/\sqrt{n}$
$\lesssim_P s \sqrt{[\log (p\vee n)]/n}$.

Step 3. (Behavior of $ii$.) Decompose
$$
ii = (m+V)'\MX (m+V)/n = V'V/n + \underset{=:ii_a}{m' \MX m/n} +
\underset{=:ii_b}{ 2 m'\MX V/n } - \underset{=:ii_c}{V'\PX V/n}.
$$
Then $|ii_a| \lesssim_P [s \log (p\vee n)]/n= o_P(1)$ by Step 4,
$|ii_b| \lesssim_P [s \log (p\vee n)]/n= o_P(1)$ by reasoning similar to deriving
the bound for $|i_b|$, and $|ii_c| \lesssim_P [s \log (p\vee n)]/n= o_P(1)$
by reasoning similar to deriving the bound for $|i_d|$.

Step 4. (Auxiliary: Bound on $\|\MX m\|$ and related quantities.)
Observe that
$$
\sqrt{ [s \log (p\vee n)]/n} \underset{(1)}{\gtrsim_P} \| \MXd m/\sqrt{n}\|
  \underset{(2)}{\gtrsim_P}  \| \MX m/\sqrt{n}\|
  \underset{(3)}{\gtrsim_P} |\| X(\tilde \beta_m(\hat I) - \beta_{m0})/\sqrt{n}\| - \| R_m/\sqrt{n}\||
$$
where inequality (1) holds  since  by Theorem 2 $\| \MXd m/\sqrt{n}\| \leq \|(X\tilde \beta_D(\hat I_1) - m)/\sqrt{n}\|
\lesssim_P \sqrt{ [s \log (p\vee n)]/n}$, (2) holds by $\hat I_1 \subseteq \hat I$, and (3) by the triangle
inequality.  Since $\|R_m/\sqrt{n}\| \lesssim \sqrt{s/n}$ by assumption ASTE, conclude
that wp $\to 1$,
\begin{eqnarray*}
\sqrt{ [s\log (p\vee n)]/n } & \gtrsim_P & \| X(\tilde \beta_m(\hat I) -
\beta_{m0})/\sqrt{n}\| \\
& \geq &  \sqrt{\semin{\hat s}} \| \tilde \beta_m(\hat I) - \beta_{m0}\|
\gtrsim_P   \| \tilde \beta_m(\hat I) - \beta_{m0}\|,
\end{eqnarray*}
since $\hat s \lesssim_P s$ by Theorem \ref{corollary3:postrate} so that $1/\semin{\hat s} \lesssim_P 1$
by condition SE.

Step 5. (Auxiliary: Bound on $\|\MX g\|$ and related quantities.)
Observe that
 \begin{eqnarray*}
\sqrt{ [s \log (p\vee n)]/n} & \underset{(1)}{\gtrsim_P} &  \| \MXy (\alpha_0 m + g)/\sqrt{n}\| \\
  & \underset{(2)}{\gtrsim_P}&  \| \MX (\alpha_0 m + g)/\sqrt{n}\|
  \underset{(3)}{\gtrsim_P} |\|  \MX g /\sqrt{n}\|- \|\MX \alpha_0 m /\sqrt{n}\||
\end{eqnarray*}
where inequality (1) holds  since  by Theorem \ref{corollary3:postrate} $\| \MXy (\alpha_0 m + g)/\sqrt{n}\| \leq \|(X \tilde \beta_{Y_1}(\hat I_2) - \alpha_0 m - g)/\sqrt{n}\|
\lesssim_P \sqrt{ [s \log (p\vee n)]/n}$, (2) holds by $\hat I_2 \subseteq \hat I$, and (3) by the triangle
inequality.  Since $\|\alpha_0\|$ is bounded uniformly in $n$ by assumption,  by Step 4,
$\|\MX \alpha_0 m /\sqrt{n}\| \lesssim_P  \sqrt{ [s \log (p\vee n)]/n}$. Hence
conclude that
$$
\sqrt{ [s \log (p\vee n)]/n}  \gtrsim_P \| \MX g /\sqrt{n}\| \geq  |\| X(\tilde \beta_g(\hat I) - \beta_{g0})/\sqrt{n}\| - \| R_g/\sqrt{n}\||
$$
where $\| R_g/\sqrt{n}\| \lesssim \sqrt{s/n}$ by condition ASTE. Then conclude similarly to Step 4 that wp $\to 1$,
$$
\sqrt{ [s\log (p\vee n)]/n } \gtrsim_P \| X(\tilde \beta_g(\hat I) -
\beta_{g0})/\sqrt{n}\| \geq \sqrt{\semin{\hat s}} \| \tilde \beta_g(\hat I) - \beta_{g0}\|
\gtrsim_P \| \tilde \beta_g(\hat I) - \beta_{g0}\|.
$$

Step 6. (Variance Estimation.)  Since $\hat s \lesssim_P s = o(n)$, $(n-\hat s - 1)/n = 1 + o_P(1)$.   Hence consider
$$
\hat \sigma^2_{\zeta} = \| (Y_1 - \check \alpha D)'\MX\|^2/n = \| (\zeta+ (\alpha_0 - \check \alpha)'D + g)' \MX \|^2/n.
$$
Then by Steps 1, 3, and 5
$$
| \hat \sigma - \| \zeta' \MX\|/\sqrt{n}\| | \leq \| g' \MX\|/\sqrt{n} + \|\check \alpha - \alpha_0\| \| D' \MX\|/\sqrt{n} \lesssim_P \sqrt{ [s \log (p\vee n)]/n } +  n^{-1/2}  = o_P(1).
$$
Moreover,
$$
\| \zeta' \MX \|^2/n = \zeta'\zeta/n - \zeta'\PX\zeta/n =  \sigma_{\zeta}^2  + o_P(1),
$$
where $\zeta'\zeta/n = \sigma_{\zeta}^2 + O_P(n^{-1/2})$ by Chebyshev inequality and
$\zeta'\PX\zeta/n  \lesssim_P [s \log (p\vee n)]/n = o_P(1)$ by the argument similar to that used to bound $|i_d|$. \qed

\bibliographystyle{econometrica}
\bibliography{mybibVOLUME}

\end{document}